\documentclass[graybox, envcountchap]{svmult}

\usepackage{mathptmx}        
\usepackage{amsmath}
\usepackage{amssymb}
\usepackage{xcolor}
\usepackage{helvet}          
\usepackage{courier}         
\usepackage{dirtree}
\usepackage[square,sort,comma,numbers]{natbib} 
\include{abb}

\usepackage{makeidx}        
\usepackage{graphicx}        
\usepackage{subfig}

\usepackage{multicol}        
\usepackage{color}
\usepackage[bottom]{footmisc}

\usepackage{hyperref}        
\hypersetup{colorlinks=true,urlcolor=blue}

\usepackage[misc]{ifsym}
\usepackage{url}

\makeindex             

\begin{document}


\title{Thin Accretion disks in GR-MHD simulations}
\titlerunning{Thin Accretion Disks} 
\author{Indu K.~Dihingia \& Christian Fendt}
\authorrunning{Dihingia \& Fendt} 
\institute{Indu K. Dihingia (\Letter) \at Tsung-Dao Lee Institute, Shanghai Jiao-Tong University, Shanghai, 520 Shengrong Road, 201210, People's Republic of China, \email{ikd4638@gmail.com, ikd4638@sjtu.edu.cn}
\and Christian Fendt \at Max Planck Institute for Astronomy, K\"onigstuhl 17, 69117 Heidelberg, Germany, \email{fendt@mpia.de}}
%
%
\maketitle

\abstract{We review some recent results of general relativistic magnetohydrodynamic (GR-MHD) simulations considering the evolution of geometrically thin disks around a central black hole. 
Thin disk GR-MHD simulations complement the widely used MAD (Magnetically Arrested Disk) or SANE (Standard And Normal Evolution) approaches of evolving from an initial disk torus.
In particular, we discuss the dynamical evolution of the disk, its role in the formation of disk winds or jets, the impact of disk resistivity, and its potential role in generating magnetic flux by an internal disk dynamo.
The main characteristics of a thin disk in our approach are the Keplerian rotation of the disk material, which allows to launch disk outflows by the Blandford-Payne magneto-centrifugal effect, in addition to the Blandford-Znajek-driven spine jet from the black hole ergosphere. Thus, for this approach, we neglect disk thermodynamics and radiative effects, concentrating predominantly on the dynamical evolution of the system.
Resistive MHD further allows the investigation of physical reconnection and also dynamo action. 
Magnetic reconnection may generate magnetic islands of plasmoids that are ejected from the disk along with the outflow.
We also discussed potential applications of thin disk in explaining the decaying phase of an outburst in black hole X-ray binaries (BH-XRBs). Post-processing of radiation using the simulated dynamical data allows to derive spectra or fluxes, e.g., in the X-ray band,
and to derive potential variability characteristics.}


\section{Introduction}
\begin{figure}
    \centering
    \includegraphics[scale=0.4]{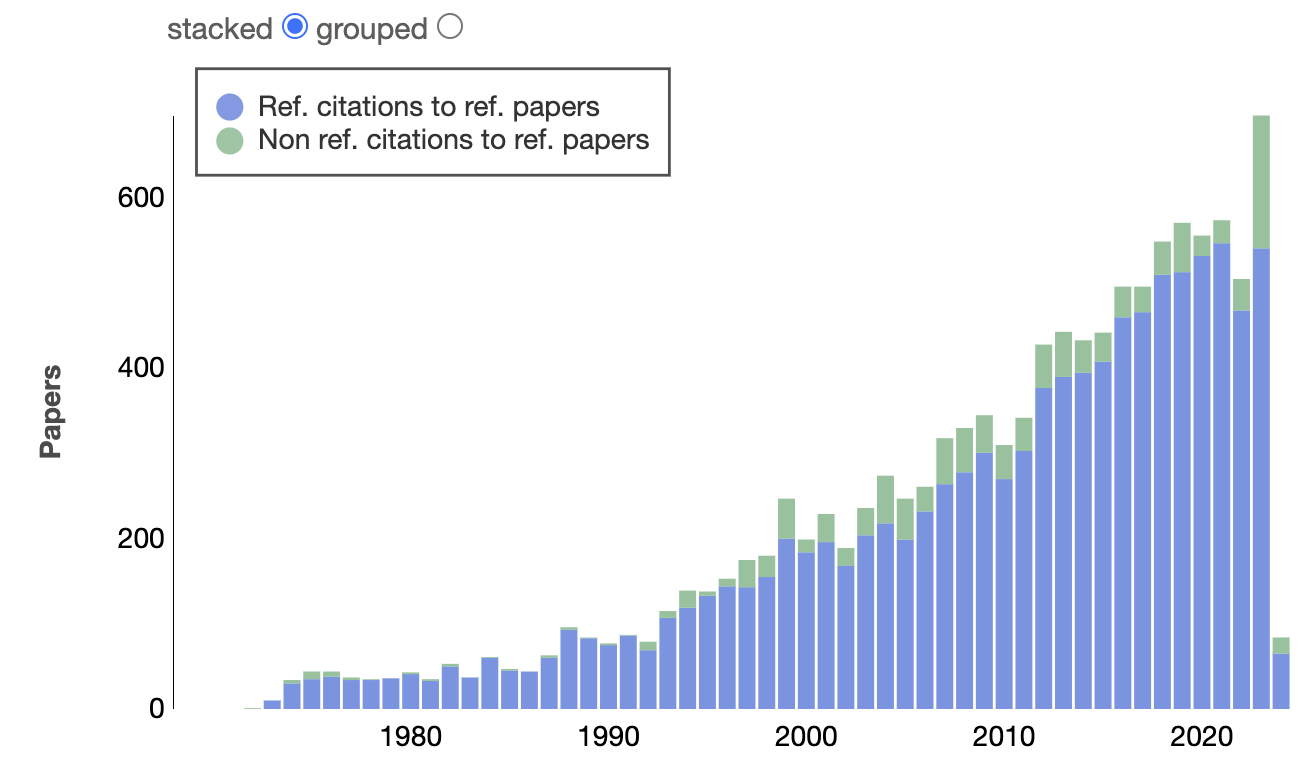}
    \caption{Citation metric for Shakura \& Sunyaev 1973 \cite{Shakura-Sunyaev1973}, accumulating 11,850 citations by March 2024.}
    \label{fig:ss1973}
\end{figure}

Thin accretion disks have been instrumental for several decades in explaining a wide range of observed astrophysical sources, including X-ray binaries (XRBs) and active galactic nuclei (AGNs).
This is evident from the enormous citation metric for the paper by Shakura \& Sunyaev from 1973 \citep{Shakura-Sunyaev1973} (SS73, see Fig.~\ref{fig:ss1973}), 
which is the milestone article for thin disks. 
SS73 deals the accretion flow in Newtonian approach, which was subsequently developed to its relativistic version by Novikov \& Thorne in 1973
\citep{Novikov-Thorne1973}. 
Thin disk models are most relevant if the astrophysical source is emitting multi-colour black body radiation (e.g., in the soft state of XRBs).  

Along with the thin disk model, many other accretion disk models have been proposed with the increase in varieties of astrophysical sources. 
For example, spherically symmetric Bondi flow \citep{Bondi:1952ni}, advection dominated accretion flows (ADAF) \citep{Narayan-Yi1995}, trans-sonic accretion disk \citep{Fukue1987, Chakrabarti1989}, ultra-luminous optically thick geometrically thick disk \citep{Abramowicz-etal1988}, two-component accretion flow (TCAF) \citep{Chakrabarti2018}, etc. However, due to the complex theoretical nature of these models, they are not widely used as the thin disk model by astronomers. With the development of cutting edge general relativistic magneto-hydrodynamic (GR-MHD) simulations, further families of accretion disks have evolved. 

Typically, GR-MHD simulations of accretion start with the initial condition of a hydrodynamically stable torus developed by Fishbone \& Moncrief 1976 \cite{Fishbone-Moncrief1976}.
Threaded by a weak magnetic field, the initial stable torus configuration becomes unstable and evolves into a disky structure. 
Depending on the initial magnetic flux, the torus evolves into a SANE disk (standard and normal evolution) or a MAD disk (magnetically arrested disk) \citep[e.g.,][]{Tchekhovskoy-etal2010,Porth-etal2021}. 
These two solutions are seen as standard solutions by many in the scientific community. 
These models (SANE and MAD) have been used to understand near horizon physics around supermassive black holes Sgr~A$^*$ and M~87$^*$.
Undoubtedly, these models adequately constrain the parameters of these two supermassive black holes, such as mass, accretion rate, inclination, and black hole spin \citep[e.g,][]{EHTI-2019,EHTVII-etal2021,EHT2022,EHT2024}. 
However, many observations of the ``soft excess'' feature found in many Seyfert galaxies (e.g., \cite{Weaver-etal1995,Boissay-etal2016,Xu-etal2021}) can be understood with the high-density reflection model (e.g., \cite{Garcia-etal2019}). Note that the ``warm corona'' model is one of the candidates to explain ``soft excess'' in Seyfert galaxies (e.g., \cite{Ballantyne:2019lqp,Mehdipour:2023hjo}). 
This is also supported by the soft X-ray reverberation signature (e.g., \cite{DeMarco-etal2013,Kara-etal2013}). 
These models essentially rely on accretion through a thin disk geometry. 
X-ray reflection models have indicated that the inner disk radius of the accretion disk is very close to the innermost stable circular orbit (ISCO) for several XRBs even in luminous hard-state \citep[see for detail][]{Liu-etal2023}. 
The caveat is that this model approach uses a ``lamp-post'' corona model, where a hot corona is considered at some vertical height from the black hole
( e.g., \cite{Matt-etal1991,Martocchia-Matt1996,Markoff-etal2005}). 
In summary, it is clear that at least some astrophysical sources (both AGNs and BH-XRBs), if not all, require a thin disk geometry around black 
holes in order to explain their observations.

Thus, the model of a thin disk evolution in GR-MHD is still appealing. We know from non-relativistic simulations that such Keplerian thin disks can easily launch a strong disk wind that can evolve into a disk jet. There is no reason to doubt that such a thin disk can also be present around (supermassive) black holes. 
With such a disk, a two-component jet structure for rotating black holes is feasible: a Blandford-Znajek (hereafter BZ; \cite{Blandford-Znajek1977}) driven Poynting flux driven high-speed jet launched from the spinning black hole and a mass loaded disk jet (driven by the Blandford-Payne magneto-centrifugal mechanism; hereafter BP; \cite{Blandford-Payne1982}) from the surrounding accretion disk. Recent GR-MHD simulations were indeed able to compare the energy output of both jet components \citep{Qian-etal2017,Qian-etal2018,Vourellis-etal2019}. 

A number of studies have been conducted considering different accretion flow models, including thin disks.
In this book chapter, we will discuss recent developments in GR-MHD simulations of jet launching and other aspects considering thin disks. 
Other chapters will provide a comprehensive understanding of the current research on various other accretion flow models and their implications 
for astrophysical phenomena. 
Additionally, we will explore the current limitations and future directions of GR-MHD simulations in understanding the dynamics of thin disk accretion flows.

\section{Basic equations}
In order to derive the magnetized flow properties, a set of GR-MHD equations need to be solved. 
The basic resistive GR-MHD equations are as follows:
\begin{align}
\begin{aligned}
&\nabla_\mu\left(\rho u^\mu\right)=0,\\
&\nabla_\mu T^{\mu\nu}=0,\\
&\nabla_\mu F^{\mu\nu}={\cal J}^\mu,\\
&\nabla_\mu{}^*F^{\mu\nu}=0,\\
\end{aligned}
\label{eq-01}
\end{align}
where different symbols have their usual meaning, such as $\rho, u^\mu, T^{\mu\nu}, F^{\mu\nu}, {}^*F^{\mu\nu}$, and ${\cal J}^{\mu}$ represent the rest-mass density, four velocity, energy-momentum tensor, Faraday tensor, dual of the Faraday tensor, and electric 4-current, respectively.
The details of these terms, as well as the explicit procedures for solving them, 
are given in \cite{Rezzolla-Zanotti2013,Porth-etal2017,Olivares-etal2019,Ripperda-etal2019,Vourellis-etal2019}. 

These equations are solved using Modified Kerr-Schild (MKS) geometry. 
By selecting an appropriate MKS stretching parameter, we ensure that the concentration of maximum resolution is near the equatorial plane of the simulation domain \citep{McKinney-Gammie2004}. 
In this chapter, we adopt a generalized unit system with $G=M=c=1$ to express any quantities. 
Here, $M$ represents the mass of the black hole, $G$ represents the universal gravitational constant, and $c$ represents the speed of light. 
The unit system employed here represents mass, length, and time using $M$, $r_g=GM/c^2$, and $t_g=GM/c^3$, respectively. 
Also, we will be following the sign convention for the metric as $(-, +, +, +)$, with four velocities satisfying $u_\mu u^\mu = -1$. 

In this chapter, we will only concentrate on thin disks around Kerr black holes. 
Accordingly, the Kerr metric in Kerr-Schild (KS) coordinates ($t, r,\theta,\phi$) is given by,
\begin{align}
\begin{aligned}
    ds^2 &= -\left(1 - \frac{2r}{\Sigma}\right)dt^2 + \left(\frac{4r}{\Sigma}\right)dr~dt + \left(1 + \frac{2r}{\Sigma}\right)dr^2 + \Sigma d\theta^2\\
    & + \sin^2\theta\bigg[\Sigma + a^2\left(1 + \frac{2r}{\Sigma}\right)\sin^2\theta\bigg]d\phi^2 - \left(\frac{4ar\sin^2\theta}{\Sigma}\right)d\phi~dt \\
    & - 2a\left(1+\frac{2r}{\Sigma}\right)\sin^2\theta d\phi~dr,
\end{aligned}
\end{align}
\citep{McKinney-Gammie2004},
where $\Sigma\equiv r^2 + a^2\cos^2\theta$ and $a$ is the rotation parameter for the Kerr black hole. The determinant of the metric will also be often used in some calculations, which is given by $g=-\Sigma^2\sin^2\theta$. Finally, the coordinate transformations from KS ($t, r,\theta,\phi$) to MKS ($x_0,x_1,x_2,x_3$) are related by,
\begin{align}
    t=x_0~; r=R_0 + e^x_1~; \theta = \pi x_2 + \frac{1}{2}h\sin(2\pi x_2)~;{\rm and} ~ \phi = x_3.
\end{align}
The parameter $h$ can be adjusted to concentrate grid zones towards the equator as it is increased from $0$ to $1$. In practice, the initial conditions are supplied to the GR-MHD codes in Boyer-Lindquist (BL) coordinates, and after proper transformations, codes solve them in MKS coordinates. Below, we show all the initial conditions in BL coordinates.

Accretion disks may also exist around other exotic objects or on non-Kerr backgrounds proposed in the literature \citep[e.g.,][and many more]{Amarilla-etal2010,Bambi-Yoshida2010,Tsukamoto-etal2014,Fathi-etal2021}. 
However, there are very few GR-MHD thin disk simulations performed on such backgrounds (e.g., \cite{Nampalliwar-etal2022}); only semi-analytical accretion solutions are available \citep[e.g.,][]{Bambi-Barausse2011,Chen-etal2012,Dihingia-etal2020,Faraji-etal2020,Patra-etal2024,Uniyal-etal2024}.  
It would be interesting to explore these alternative scenarios in future studies to understand the behaviour of accretion disks in diverse 
gravitational environments. 
Such investigations could provide valuable insights into the dynamics and observational signatures of accretion processes beyond Kerr black holes.

\section{The thin disk prescription}
GR-MHD simulations rely on the initial conditions, such as initial density, velocities, magnetic field components, resistivity profiles, etc. In this section, we give the basic profiles of these initial conditions. Generally, for density and velocities, there are two broad ways of setting them: a (i) Shakura-Sunyaev-like disk model \cite{Shakura-Sunyaev1973}, or a (ii) Novikov-Thorne relativistic disk model \cite{Novikov-Thorne1973}. In the following, we briefly discuss them explicitly, along with other initial conditions.

\subsection{The Shakura-Sunyaev disk (SSD)}
One option to prescribe an initially thin disk is to apply a density and pressure distribution typically used in non-relativistic MHD simulations 
of jet launching \cite{CasseKeppens2002, Zanni-etal2007, Sheikhnezami-etal2012, StepanovsFendt2016}. 
This disk setup has been proven to be stable in the hydrodynamical limit and evolves smoothly for the MHD case, in which strong disk winds can be launched
that may turn into collimated jets.
The idea for applying this setup to the GR-MHD approach is that relativistic deviations are minor for most parts of the disk, assuming an inner disk radius outside the ISCO. After all, there is a dynamical evolution of the disk-outflow system that will settle into a new dynamical equilibrium considering the effects of accretion, outflow, and relativity.

In the simulations described below, an initial disk density distribution is applied as described by a non-relativistic vertical equilibrium profile, such as applied in \citep{Zanni-etal2007,Sheikhnezami-etal2012, Qian-etal2017, Vourellis-etal2019}, slightly modified to fit into the GR-MHD code,
\begin{equation}
    \rho(r,\theta) = \left[ \frac{\Gamma -1}{\Gamma} \frac{r_{\rm in}}{r} \frac{1}{\epsilon^2} \left( \sin \theta + \epsilon^2 \frac{\Gamma}{\Gamma -1} \right) \right]^{1 / (\Gamma -1)},
    \label{eq-ssd-disk}
\end{equation}
Here, $r_{\rm in}$ is the initial inner disk, and $\epsilon = H/r$ is the initial disk aspect ratio as defined by the vertical equilibrium of a disk with a local pressure scale height $H(r)$. The pressure and internal energy are given by the polytropic equation of state $p = K \rho^{\Gamma}$, where $K$ is the polytropic constant, and the internal energy $u = p /(\Gamma -1)$.

Conveniently, for the initial orbital velocity a profile following \cite{Paczynsky1980} can be applied,
\begin{equation}
{\tilde{u}}^{\phi } = {r}^{-3/2} \left(\displaystyle \frac{r}{r-{R}_{\mathrm{PW}}}\right),
\end{equation}
where the smoothing length scale for the gravity $R_{\mathrm PW}$ is a constant of choice, for example chosen to be equal to the gravitational radius. This approximation is applied in the $\phi$-component of the fluid velocity $u^{\phi}$.

The initial velocity profile of the disk will not be in equilibrium with the underlying general relativistic gravity around the black hole. 
However, with temporal evolution, the disk rotational velocity will soon approach a new dynamical equilibrium, which depends on governing GR-MHD variables such as gas pressure and Lorentz forces.
\subsection{The Novikov-Thorne disk (NTD)}
In thin disk approximation, most of the matter occupies the equatorial plane of the coordinate system ($\theta=\pi/2$) and the $\theta$  component of the four-velocity $u^\mu$ can be neglected $(i.e., u^\theta = 0)$. With these approximations, the accretion rate $(\dot{M_0})$, time average radiation flux $(F)$, and time average torque $(W^r_\phi)$ are obtained as \cite{Novikov-Thorne1973,Page-Thorne1974},
\begin{align}
\begin{aligned}
\dot{M}_0&=-2\pi r \Sigma u^r,\\
F&=\left(\dot{M}_0/4\pi r\right)f,\\
W^r_\phi&=\left(\dot{M}_0/2\pi r\right)\left[(E^\dagger-\Omega L^\dagger)/(-\Omega_{,r})\right]f,\\
\end{aligned}
\label{eq-02}
\end{align}
where $E^\dagger = -u_{t,e}$, $L^\dagger = u_{\phi,e}$, and $\Omega = u^{\phi}_e/u^{t}_e$. The subscript $e$ represents quantities defined on the equatorial plane. Considering the flow motion to be Keplerian $(u^r\sim0)$, i.e., rotation supported accretion disk, the explicit expression of $f$ is obtained as
\begin{align}
f=-\Omega_{,r}(E^\dagger-\Omega L^\dagger)^{-2}\int_{r_{ms}}^r (E^\dagger-\Omega L^\dagger)L^\dagger_{,r}dr.
\label{eq-03}
\end{align}
Here, $r_{ms}$ is the marginally stable radius or the ISCO radius: (for Kerr black hole it is the radius at which $\frac{dE^\dagger}{dr}=\frac{dL^\dagger}{dr}=0$). The explicit expression of $f$ is given by

\begin{align}
\begin{aligned}
f =& \frac{3}{2} \frac{1}{ x^2 \left(2 a+x^3-3 x\right)}\bigg[ x - x_0 -\frac{3}{2}\ln\left(\frac{x}{x_0}\right) \\
&- \frac{3\left(s_1-a\right)^2}{s_1(s_1-s_2)(s_1-s_3)} \ln \left(\frac{x- s_1}{x_0-s_1}\right) \\
&- \frac{3\left(s_2-a\right)^2}{s_2(s_2-s_1)(s_2-s_3)}\ln \left(\frac{x- s_2}{x_0-s_2}\right) \\
&- \frac{3\left(s_3-a\right)^2}{s_3(s_3-s_1)(s_3-s_2)}\ln \left(\frac{x- s_3}{x_0-s_3}\right)\bigg], \\
\end{aligned}
\label{eq-04}
\end{align}
where we use $x=\sqrt{r}$, implying $x_0=r_{ms}^{1/2}$, and $s_1,s_2$ and $s_3$ are the three real roots of $s^3 - 3s + 2a=0$, which are given bellow,
\begin{align}
\begin{aligned}
s_1=&2 \cos \left(\frac{1}{3} \cos ^{-1}(a)-\frac{\pi }{3}\right),\\
s_2=&2 \cos \left(\frac{1}{3} \cos ^{-1}(a)+\frac{\pi }{3}\right),\\
s_3=&-2 \cos \left(\frac{1}{3} \cos ^{-1}(a)\right).\\
\end{aligned}
\label{eq-05}
\end{align} 
By considering the flux to be emitted as a multi-color black body, we can calculate the temperature of the thin disk. The explicit form of the temperature is given by:
\begin{align}
\frac{p_e}{\rho_e} \propto T_{\rm bb}(x)\propto F^{1/4} = \Theta_0 \left(\frac{f(x)}{x^2}\right)^{1/4},
\label{eq-06}
\end{align}
where $p_e$ and $\rho_e$ are the pressure and density at the equatorial plane. $\Theta_0$ is a constant related to the initial temperature of the disk. With the polytropic equation of state $p = {\cal K}\rho^\Gamma$, the density of the fluid at the equatorial plane is expressed as,
\begin{align}
\rho_e=\left(\frac{\Theta_0}{\cal K}\right)^{1/(\Gamma -1)} 
              \left(\frac{f(x)}{x^2}\right)^{1/(4(\Gamma - 1))},
\label{eq-07} 
\end{align}
where ${\cal K}$ is a constant associated with the flow's entropy. The density profile at the equatorial plane ($\theta=\pi/2$) is provided by Eq. (\ref{eq-07}). The general density profile for the off-equatorial plane can be formulated as follows:
\begin{align}
\rho(r,\theta) = \rho_e \exp\left(-\frac{\alpha^2 z^2}{H^2}\right); ~~ z=r\cos(\theta).
\label{eq-08}
\end{align}
Equation (\ref{eq-08}) is written considering density drops along the vertical direction that follow a normal distribution.
In order to ensure a thin disk geometry, we choose $\alpha=2$. 
Here, $H$ is the scale height of the initial accretion disk. 
We follow Riffert \& Herold 1995 \cite{Riffert-Herold1995} and Peitz \& Appl 1997 \cite{Peitz-Appl1997} to obtain $H$,
\begin{align}
H^2 = \frac{p_e r^3}{\rho_e {\cal F}},
\label{eq-09}
\end{align}
where
$$
{\cal F}=\gamma_\phi^2\frac{\left(a^2+r^2\right)^2+2 a^2 \Delta }{\left(a^2+r^2\right)^2-2 a^2 \Delta },
$$
with $\gamma_\phi^2 = \left(1-\Omega\lambda\right)^{-1}$ and $\Delta = r^2-2 r + a^2$ where $\lambda = - u_{\phi, e}/u_{t,e}$.  

Along with the density profile, we need to supply the initial four velocities ($u^\mu$). 
To calculate $u^\phi$, we consider that the fluid element follows the geodesic equation,
\begin{align}
u^t u^t \Gamma^r_{tt}+ 2 u^t u^\phi \Gamma^r_{t\phi}+ u^\phi u^\phi \Gamma^r_{\phi\phi}=0.
\label{eq-10}
\end{align}
Also, the four velocities must satisfy
\begin{align}
g_{\mu\nu}u^\mu u^\nu = g_{\phi\phi}u^\phi u^\phi + g_{tt}u^t u^t + 2 g_{t\phi}u^t u^\phi = -1.
\label{eq-11}
\end{align}
Here, $g_{\mu\nu}$ is the metric tensor, and $\Gamma^\alpha_{\beta \gamma}$ are the Christoffel symbols. Solving Eqs. (\ref{eq-10}) and (\ref{eq-11}), we obtain the expressions of $u^\phi$ and $u^t$. The explicit expression of  $u^\phi$ is given by
\begin{align}
u^\phi(r,\theta) = \left(\frac{\cal A}{{\cal B}+ 2 {\cal C}^{1/2}}\right)^{1/2},
\label{eq-12}
\end{align}
where
$$
\begin{aligned}
{\cal A}=&\left(\Gamma^r_{tt}\right)^2,\\
{\cal B}=&g_{tt}\left(\Gamma^r_{tt}\Gamma^r_{\phi \phi}-2 {\Gamma^r_{t\phi}}^2\right)+2 g_{t\phi} \Gamma^r_{tt} \Gamma^r_{t\phi} - g_{\phi \phi } {\Gamma^r_{tt}}^2,\\
{\cal C}=&\left({\Gamma^r_{t\phi}}^2 - \Gamma^r_{tt} \Gamma^r_{\phi \phi}\right) (g_{t\phi} \Gamma^r_{tt}- g_{tt} \Gamma^r_{t\phi})^2.\\
\end{aligned}
$$
The explicit expression of $u^t (r,\theta)$ can be calculated by substituting $u^\phi (r,\theta)$ to any of the equations, 
either Eq.~(~\ref{eq-10}) or Eq.~(\ref{eq-11}). 
Thus, the expressions of $u^\phi$ and $u^t$ on the equatorial plane ($\theta=\pi/2$) are given by
\begin{align}
\begin{aligned}
u^\phi_e =& \frac{1}{\sqrt{x^3 \left(2 a+x^3-3 x\right)}},\\
u^t_e= & \frac{a+x^3}{\sqrt{x^3 \left(2 a+x^3-3 x\right)}}.\\
\end{aligned}
\label{eq-13}
\end{align}
After all, $\rho(r,\theta)$ and $u^\phi(r,\theta)$ obtained in Eq. (\ref{eq-08}) and Eq. (\ref{eq-12}) serve as initial conditions 
for the thin disk for simulation models.

\subsection{The initial magnetic field}
Along with the velocities and density of a rotation-supported thin disk, one needs to supply initial magnetic fields that trigger magneto-rational instabilities (MRI). MRI transports angular momentum \citep{Balbus-Hawley1998}, thereby setting accretion in the system. Moreover, MRI aids in the development of large-scale magnetic fields, which are responsible for launching relativistic jets and winds. Typically, in GR-MHD codes, a vector potential for inclined large-scale magnetic fields threading the disk is supplied. One such vector potential, involving non-zero $\phi$ components, is described by \cite{Zanni-etal2007,Vourellis-etal2019}.
\begin{align}
A_\phi \propto \left(r \sin \theta\right)^{3/4} \frac{m^{5/4}}{\left(m^2 + (\tan\theta)^{-2}\right)^{5/8}}.
\label{eq-vec1}
\end{align}
The parameter $m$ is related to the initial inclination of the field lines and determines the system's magnetic flux. The parameter $m$ is crucial for launching BP
type wind from the accretion disk \citep{Blandford-Payne1982,Dihingia-etal2021}. The strength of the poloidal magnetic field depends on the initial supplied minimum plasma-$\beta$ parameter $\beta_{\rm min}$.

\subsection{Resistivity and magnetic diffusivity}
The major breakthrough of the thin disk model by Shakura and Sunyaev came with their introduction of a {\em turbulent viscosity} $\nu$ as the major agent for angular momentum transport in accretion disks. This was accompanied by viscous heating that can result in the observed luminosities in stellar compact binaries or quasars.

This turbulent viscosity is understood as an {\em anomalous} viscosity, thus much stronger than the microscopic molecular viscosity. At that time, the cause of the turbulence was yet unknown\footnote{It is now accepted that it is caused by the MRI \citep{BalbusHawley1991}.}. Fundamentally, \cite{Shakura-Sunyaev1973} introduced the terminology of an $\alpha$-viscosity, which can be approximated in strength by the typical sound speed $c_S$ in the disk and a typical length scale $H$ (the disk scale height), resulting in values of $\alpha < 1$ that determine the strength of the viscous stress and the disk angular momentum transport, involving the turbulent viscosity $ \nu = \alpha c_s H$, and involving the tangential stress between two disk layers, $w_{r,\phi} = - \alpha \rho c_s^2$.

Similar to the turbulent viscosity, one may infer a turbulent resistivity or, respectively, a turbulent magnetic diffusivity $\eta$. The interpretation is that turbulent motions in the disk allow the material to diffuse across the magnetic field. In particular, this allows for mass loading onto the disk wind from the accreting material. Connected to turbulent diffusivity and resistivity, there are two major physical effects that play roles in disk-jet evolution. One is the Ohmic heating of the disk gas, and the other is the reconnection of the magnetic field.

In non-relativistic MHD, the resistive term enters the system of equations via the induction equation for the field evolution and the energy equation for the Ohmic heating. In general relativity, the situation is more complex as resistivity enters via Ohm's law and the prescription of the electric field,
\begin{equation}
e^{\mu} = \eta j^{\mu}
\end{equation}
with the electric four vector $e^{\mu}$ and the electric current density in the comoving observer's frame 
(in contrast to ${\cal J^\mu}$ as seen by the normal observer).
The electric field observed by a normal observer, split into the usual 3+1 form is $\cal{E}^{\mu}$. 
\cite{Bucciantini-etal2013} have worked out a closure relation for general relativistic resistive MHD that treats the time evolution of the electric field four vector $\mathcal{E}^{\mu}$ as
\begin{equation}
\gamma^{-1/2} \partial_t\left(\gamma^{1/2} \vec{\mathcal{E}}\right) 
- \nabla \times \left(\alpha \vec{\mathcal{B}} - \vec{\beta} \times \vec{\mathcal{E}} \right) 
+ \left(\alpha \vec{v} - \vec{\beta} \right)
= -\frac{\alpha \Gamma}{\eta} \left[ \vec{\mathcal{E}} + \vec{v} \times \vec{\mathcal{B}} - \left( \vec{\mathcal{E}} \cdot \vec{v}\right)\vec{v}\right]
\label{eqn-resE}
\end{equation}
with a Lorentz factor $\Gamma$, the determinant of its spatial 3-metric $\gamma = \sqrt{-g} /\alpha$, while $\vec{\beta} = { \beta^i }$ is the spatial shift vector in the 3+1 formalism. We refer to \cite{Bucciantini-etal2013, Qian-etal2017, Vourellis-etal2019} for a detailed derivation. Note the $\eta$ in the denominator on the r.h.s which makes the resistive treatment stiff for low diffusivity.

For alternative implementations of resistivity in (general) relativistic MHD codes and further applications, do refer to \citep{Komissarov2007,DumbserZanotti2009,Palenzuela-etal2009,Dionysopoulou-etal2013,Bugli-etal2014,Dionysopoulou-etal2015,Ripperda-etal2019,Mignone-etal2019,Ripperda-etal2020}

\subsection{A mean-field disk dynamo}
An interesting property of the closure relations for relativistic resistivity discussed above is that, in addition to the diffusive (loss) term for the magnetic field, a source term can also be added in a similar fashion. 
That provides the option to implement a mean-field dynamo term \cite{Bucciantini-etal2013,VourellisFendt2021}, similar to previous non-relativistic studies \cite{Stepanovs-etal2014, Zhou2024}, that apply the mean-field dynamo theory \cite{Steenbeck1969, Krause1980, Pudritz1981}.

Here, the mean-field usual $\alpha$-dynamo parameter is replaced by $\xi \equiv - \alpha$ to avoid confusion with the gravitational lapse. 
This parameter measures the strength of the dynamo and is thought to result from a mean effect of disk turbulence.
Consequently, when turbulence decays, the dynamo effect becomes weaker.
This is also understood as dynamo quenching.

In covariant form and in the fluid frame, Ohm's law is written as
\begin{equation}
{e}^{\mu }= \eta {j}^{\mu }\,+\,\xi {b}^{\mu },
\end{equation}
where $j^{\mu}$ is again the four vector of the electric current density. Essentially, now the electric field can no longer be calculated by the cross-product of fluid velocity and magnetic field, and new equations need to be formulated for the evolution of the electric field,
\begin{eqnarray}&&\begin{array}{l}{\gamma }^{-1/2}{\partial }_{t}\left({\gamma }^{1/2}{{ \mathcal E }}^{i}\right)-{\epsilon }^{{ijk}}{\partial }_{j}\left(\alpha {{ \mathcal B }}_{k}-{\epsilon }_{{knm}}{{ \mathcal E }}^{m}{\beta }^{n}\right)\\ \quad =\,-\alpha {\rm{\Gamma }}\left[{{ \mathcal E }}^{i}+{\epsilon }^{{ijk}}{v}_{j}{{ \mathcal B }}_{k}-({{ \mathcal E }}^{k}{v}_{k}){v}^{i}\right]/\eta \\ \quad +\alpha \xi {\rm{\Gamma }}\left[{{ \mathcal B }}^{i}-{\epsilon }^{{ijk}}{v}_{j}{{ \mathcal E }}_{k}-({{ \mathcal B }}^{k}{v}_{k}){v}^{i}\right]/\eta \\ \quad -q(\alpha {v}^{i}-{\beta }^{i}).\end{array}\end{eqnarray}
(Note the difference to Eqn.~\ref{eqn-resE}). For a more detailed derivation and tests of the dynamo rHARM3D code we refer to \cite{Bucciantini-etal2013,VourellisFendt2021}.

\begin{figure}
    \centering
    \includegraphics[scale=0.60]{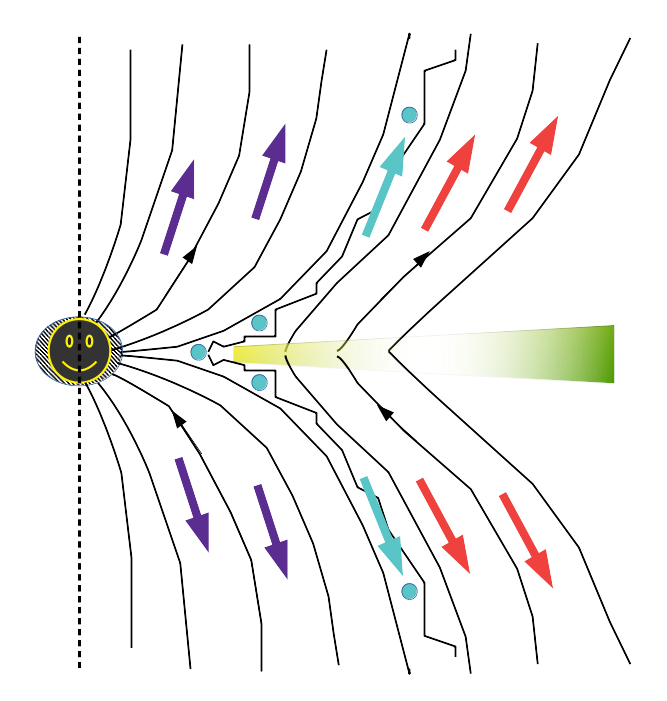}
    \caption{Schematic diagram of the disk-jet-wind configuration.
    A spinning black hole (black circle, rotational axis dashed) is surrounded by a thin disk in Keplerian rotation (greenish block).
    Magnetic field lines (black lines) that are rooted in the ergosphere (hatched ellipse) launch a BZ jet (purple arrows).
    Magnetic field lines rooted in the disk launch a BP disk wind (red arrows).
    The anti-aligned magnetic field in the plunging region leads to reconnection, magnetic islands and plasmoids (turquoise circles) that are ejected along a boundary layer (jittered field line) between disk wind and spine jet. Note the upward magnetic field direction (black arrows).}
    \label{fig:cartoon}
\end{figure}

\section{Characteristics of the black hole-disk-jet structure}
In Fig.~\ref{fig:cartoon}, we present a schematic diagram  of the disk-jet structure surrounding a central black hole as based on our simulation results. We observe such a structure in our reference model developing after several tens of inner disk revolutions and see it remaining like that up to the end of the simulation, corresponding to
several hundreds of inner disk revolutions. 

The high-density thin accretion disk occupies the area near the disk equatorial plane. The region near the rotation axis is occupied by very low density, relativistic, and highly magnetized BZ-jets, where the ratio $B_{\rm tor}/B_{\rm p}\ll 1$. The shape of the BZ-jet is supported by a region with $B_{\rm tor}/B_{\rm p}\gg 1$. We called this region as a $B_{\rm tor}$ dominated disk wind, and the driving mechanism for this disk wind is the excess toroidal magnetic pressure. $B_{\rm tor}$ dominated disk wind is turbulent due to the different instabilities present in the thin accretion disk, and we observe plasmoids in this region. Further, such a disk wind could play a very important role in the collimation or the de-collimation of the BZ-jet due to external support.

The $B_{\rm tor}$ dominated disk wind region is a potential site for turbulent mass loading from the accretion disk \cite{Britzen-etal2017}. The area between the $B_{\rm tor}$ dominated wind and the accretion disk is occupied by BP disk wind. In this region, the flow around the disk surface has $B_{\rm tor}/B_{\rm p}\lesssim 1$, the driving mechanism of the BP disk wind is the magneto-centrifugal acceleration as proposed by \cite{Blandford-Payne1982}. The flow leaves the surface of the accretion disk with sub-Alfv\'enic poloidal velocity ($M_{\rm A, p}<1$). As the BP disk wind moves far from the equatorial plane, the ratio $B_{\rm tor}/B_{\rm p}$ increases, eventually the ratio becomes $B_{\rm tor}/B_{\rm p}\gg 1$ and the flow attains a super-Alfv\'enic poloidal velocity ($M_{\rm A, p}>1$). Thus, both $B_{\rm tor}$ dominated disk wind and BP disk wind are super-Alfv\'enic and $B_{\rm tor}/B_{\rm p}\gg 1$ far from the equatorial plane, in such scenario, the flow self-collimates and evolves into a large-scale jet \cite[and references therein]{Fendt2006}.

\begin{figure}
    \centering
    \includegraphics[scale=0.40]{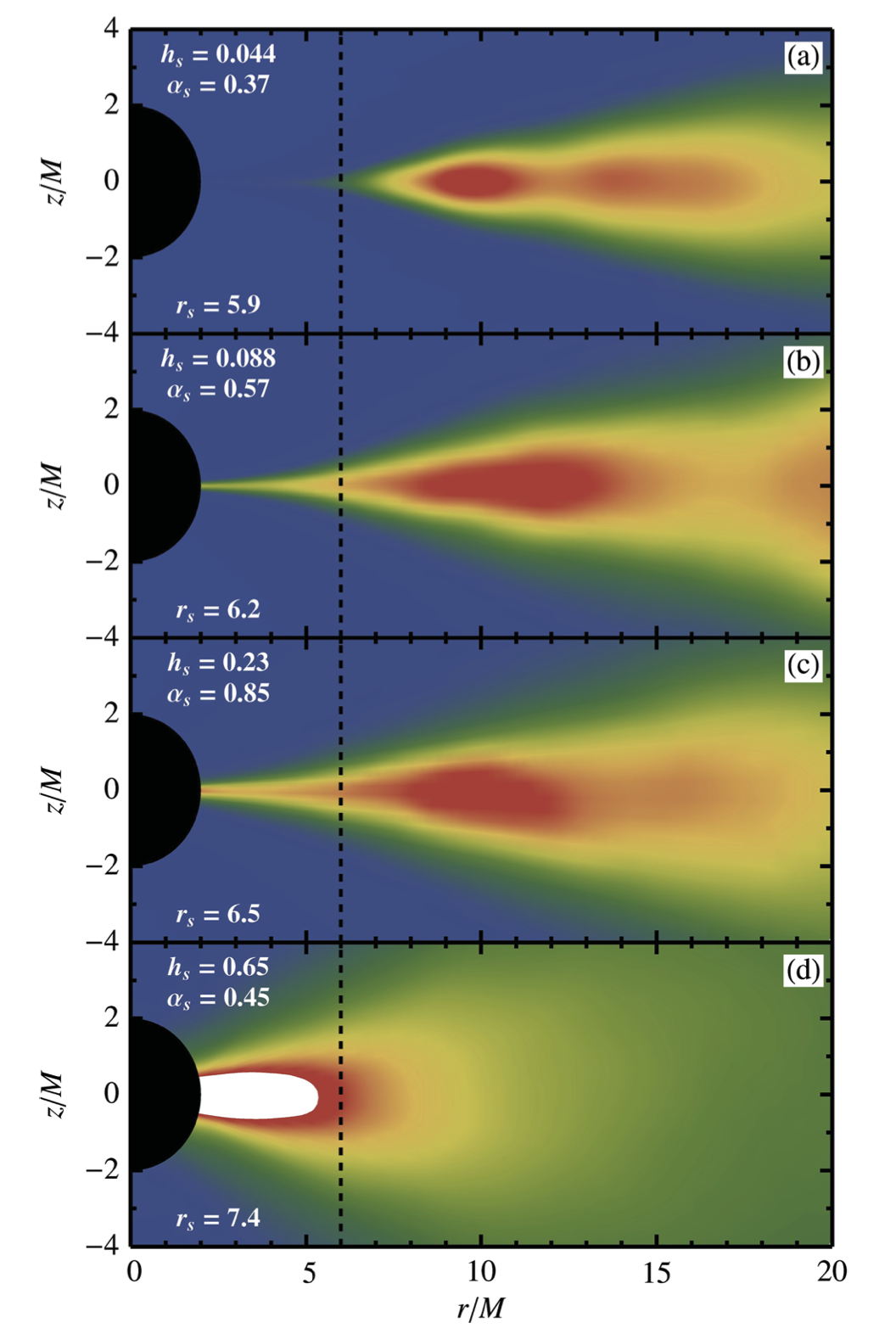}
    \caption{Time-averaged density distribution in the $r-z$ plane for four GR-MHD simulations with various disk thicknesses around a non-rotating black hole ($a=0$). The dashed vertical line shows the radial location of the ISCO. The disk aspect ratio, $h_s = H/r$, and effective Shakura–Sunyaev viscosity, $\alpha$, are marked on each panel. This figure is reproduced from \cite{Penna-etal2012}, with kind permission of Robert F.~Penna.}
    \label{fig-penna}
\end{figure}

\section{Transition from thick to thin disks}
In the two thin disk models (SSD and NTD), an initial thin disk is considered, and its evolution can be studied \citep[e.g.,][]{Qian-etal2017,Vourellis-etal2019,Dihingia-etal2021,Liska-etal2022}.
However, it is known from the basic understanding of the accretion process that for cooling dominated flow, a geometrically thick flow should be transformed into a geometrically thin flow with time.
Many studies of thin disks follow this idea and supply a rotation supported initial torus (e.g., FM torus \cite{Fishbone-Moncrief1976}) and evolve it in the presence of radiation/cooling \citep[][etc.]{Noble-etal2009,Noble-etal2010,Noble-etal2011,Schnittman-etal2016,Shashank-etal2022,Nampalliwar-etal2022}. 
To radiate the majority of heat, these models use a cooling function to keep the gas temperature at a fraction of the virial temperature. 
This allows for control over the disk's aspect ratio $(H/r)$, which is a crucial parameter in these models. 
By considering the emissivity of a thin disk, the target temperature $(T_*)$ is expressed as,
\begin{align}
    T_*=\frac{\pi}{2}\frac{R_z(r)}{r}\bigg[ \frac{H(r)}{r}\bigg]^2.
\end{align}
where $R_z(r)$ is the correction due to the vertical component of gravity \cite{Abramowicz-etal1997,Krolik1999}. 
In this approach, the thickness of the disk is maintained  by introducing cooling bound portions where the local temperature is greater than that of the target temperature $T_*(r)$. The bound matter satisfies $-hu_t<1$ and $p/\rho>T_*$, where $h$ is the specific enthalpy of the flow. 
With this, a cooling function is introduced in the  stress-energy conservation equation, i.e., $\nabla_\mu T^{\mu\nu}=-{\cal L}u^\nu$ (e.g., Eq. (17) of Noble et. al. 2009 \cite{Noble-etal2009}).

This model has been used to establish the NTD model by making one-to-one comparisons between simulations and analytical expectations. Early simulations showed differences from the NTD model at the ISCO of about $\sim10\%$ \cite{Hawley-etal2002,Noble-etal2009}. Other studies \cite{Penna-etal2010} found smaller differences of only $<3\%$. Penna et al. (2010,2012) \cite{Penna-etal2010,Penna-etal2012} analyzed potential causes for the differences and proposed that the findings are actually more aligned than they initially seemed. Fig.~\ref{fig-penna} (Figure 1 of \cite{Penna-etal2012}) shows time-average density distributions for four different GR-MHD runs for different disk heights and with $a=0$. These panels suggest the formation of NTD for aspect ratio $h_s\ll\alpha$ (viscosity parameter). They concluded that the NTD model aligns closely with numerical results quite well \cite{Penna-etal2010,Penna-etal2012}. 
These findings give leverage to directly use NTD or SSD as the initial condition for performing GR-MHD simulations \cite{Abramowicz-Fragile2013}. Rather than performing elaborate simulations, researchers can save time and resources by utilizing the NTD or SSD models as initial conditions. This allows for more efficient and accurate simulations of astrophysical phenomena. 

\section{Temporal evolution of the disk}
\subsection{Hydrodynamic evolution}
\begin{figure}
    \centering
    \includegraphics[scale=0.40]{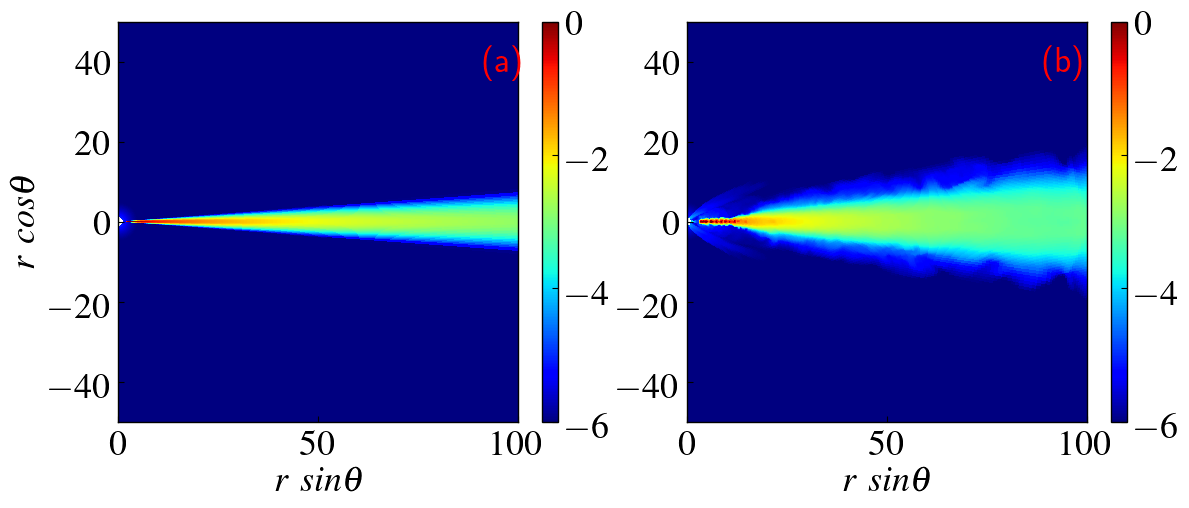}
    \caption{Hydrodynamic evolution of NTD setup from \cite{Dihingia-etal2021}. Panel (a) and (b) show density distribution at time $t=0$ and $t=10000\,t_g$, respectively.}
    \label{fig-hyd}
\end{figure}

The initial setup (irrespective of SSD or NTD) is expected to be in rotation supported hydrodynamic equilibrium, provided the aspect ratio of the accretion disk is $H/r\ll1$. To test this, the evolution of a hydrodynamical NTD asymmetric setup with $A_{r,\theta,\phi}=0$ is shown (for detailed parameters, see \cite{Dihingia-etal2021}) in Fig. \ref{fig-hyd}. The figure shows the density distribution of the setup at simulation times $t=0$ and $t=10000$ in panels Fig.~\ref{fig-hyd}a and \ref{fig-hyd}b, respectively. 
We note that the thin disk structure experiences minimal alterations over time, particularly in the outer region of the disk. It should be emphasized that the hydrodynamical model examined does not include any explicit effects of viscous heating or radiative cooling. 
Indeed, radiative cooling is essential for preserving the thin disk structure of the accretion disk. 
Nevertheless, the local heat balance condition of the thin disk ensures that the combined impact of cooling and explicit viscous heating is insignificant (refer to \cite{Shakura-Sunyaev1973, Paczynski-Kogan1981}, etc.). Consequently, it does not influence the dynamic progression of the NTD setup. The setup or similar setups (e.g., SSD) are well-suited for investigating the impact of the magnetic field on the dynamic characteristics of the accretion disk, as well as the launch of the jet and the driving of the disk wind, even under the ideal GR-MHD conditions.

\subsection{Magnetohydrodynamic evolution} 
Recently, Dihingia et. al. (2021) and Dihingia \& Vaidya (2022) \cite{Dihingia-etal2021, Dihingia-Vaidya2022a} have done an extensive parameter-space survey of axisymmetric thin accretion disks, considering different values of magnetic field strengths, inclination of field lines, and disk aspect ratio ($H/r$).

\begin{figure}
    \centering
    \includegraphics[scale=0.32]{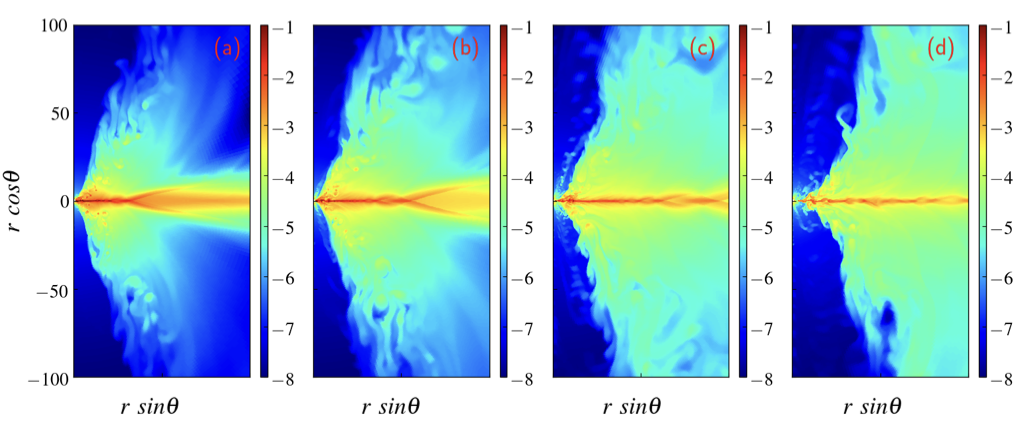}
    \caption{Magneto-hydrodynamic evolution of NTD setup from \cite{Dihingia-etal2021}. Panel (a), (b), (c), and (d) show density distribution at time $t=500\,t_g$, $1000\,t_g$, $2000\,t_g$, and $t=4000\,t_g$, respectively.}
    \label{fig-mhd-ev}
\end{figure}

In Fig.\ref{fig-mhd-ev}, the time evolution of a thin disk in the presence of an inclined magnetic field (Eq.\ref{eq-vec1}) from \cite{Dihingia-etal2021} is depicted. Following the evolution, the accretion process transitions to accretion-ejection and a disk-jet, where three major components can be identified.
\begin{enumerate}
   \item  Low density funnel near the black hole rotation axis.
   \item  The off-equatorial disk wind part has a lower density than the accretion disk part.
   \item Thin, high-density accretion disk near the equatorial plane.
\end{enumerate}

Note that relativistic jets launched due to the BZ mechanism can be seen in the low density funnel region. 
This is commonly seen in magnetized accretion flow, irrespective of the geometrically thin or thick disk \citep{Davis-Tchekhovskoy2020,Mizuno2022}. 
Other chapters from the books will describe its properties in detail; we avoid doing it here to prevent repetitions. 
Next, we discuss phenomena that are more specific to a thin disk geometry.

\begin{figure}
    \centering
    \includegraphics[scale=0.32]{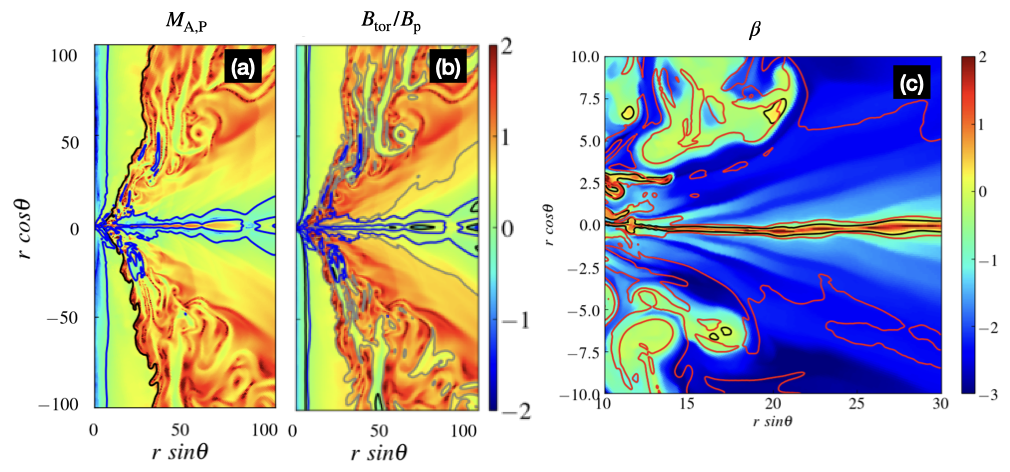}
    \caption{Distribution of (a) poloidal Alfv\'enic Mach number ($M_{\rm A,P}$), (b) $B_{\rm tor}/B_{\rm p}$, and (c) plasma-$\beta$ for a magnetised thin accretion disk evolution from \cite{Dihingia-etal2021}. In panel (a), the solid black and blue lines correspond to contours of magnetisation $\sigma=1$ and $M_{\rm A,P}=1$, respectively. In panel (b),  the solid black, grey, and blue lines correspond to contours of $B_{\rm tor}/B_{\rm p} = 1, 5$, and $M_{\rm A,P} = 1$, respectively. In panel (c), the black and red lines correspond to the contours $\beta=1$, and $M_{\rm A,P} = 1$, respectively.}
    \label{fig-wind-prop}
\end{figure}

\cite{Dihingia-etal2021} have also discussed the launching of disk winds from the a thin accretion disk in both channels, i.e.,
$B_{\rm tor}$ dominated and BP driven. 
Figure~\ref{fig-wind-prop} shows a collection of some representative results. 
Panel Fig.~\ref{fig-wind-prop}(a) shows distribution of the Alfv\'enic poloidal Mach number ($M_{\rm A,P}$), where black and blue contours correspond to $\sigma=b^2/\rho=1$ and $M_{\rm A,P}=1$, respectively, for a typical simulation model. Panel Fig.~\ref{fig-wind-prop}(b) shows the distribution of the ratio of toroidal to poloidal magnetic field strength ($B_{\rm tor}^2/B_{\rm p}^2$), where the solid black, grey, and blue lines correspond to contours of $B_{\rm tor}/B_{\rm p} = 1, 5$, and $M_{\rm A,P} = 1$, respectively. 

These two figures suggest that in this model for radii $r\gtrsim10$, the disk wind leaves the disk surface with sub-Alfv\'enic velocity and becomes super-Alfv\'enic after travelling a certain distance along the vertical direction. Within the high-density region, we observe that the flow is super-Alfv\'enic $M_{\rm A,P} > 1$ and the ratio $B_{\rm tor}/B_{\rm p} \sim 1$. Also, noticed that in the region with sub-Alfv\'enic ($M_{\rm A,P} < 1$) poloidal velocity, the ratio $B_{\rm tor}/B_{\rm p}$ is of the order of unity ($\gtrsim5$). Finally, by looking at the plasma-$\beta$ distribution in Fig.~\ref{fig-wind-prop}(c), we observe that the flow on the surface of the disk is magnetically dominated ($\beta<1$, see the black contour), where the black and red lines correspond to the contours $\beta=1$, and $M_{\rm A,P}  = 1$, respectively. These observations strongly suggest that the matter is magnetically driven in the form of a wind. This process shows similarities with the properties of disk winds, as suggested by Blandford \& Payne (1982) \cite{Blandford-Payne1982}. 

The strength of the magnetic field has a large impact on how jets and disk winds evolve and become accelerated. 
For runs with stronger magnetic fields, there is evidence of a wider jet at the base that moves with a terminal Lorentz factor that is about $\gamma\sim10$. 
The contribution of winds driven by BP mechanism goes up, while the contribution of winds driven by $B_{\rm tor}$ goes down with magnetic field strengths. 
Not only does the strength of the magnetic field matter, but the structure of the magnetic field also has a large effect on the disk, jet, and disk-wind structures. 
It seems that runs with a higher magnetic field inclination (lower value of $m$) and a higher magnetic flux around the disk surface have a wider jet with a higher Lorentz factor than runs with more vertical magnetic field lines at the disk surface. For low inclined magnetic fields, the wind is most likely be driven by BP mechanism. In cases with vertical field lines, extra toroidal magnetic pressure is more likely to be the main factor that launches the disk wind, i.e., $B_{\rm tor}$ driven disk wind.

In the reference run of Dihingia et al. (2021) \cite{Dihingia-etal2021} with $m=0.1$ and $\beta_{\rm max}\sim188$, featuring a MAD, the appearance of plasmoids due to reconnection close to the black hole is observed. Toroidal magnetic fields carry these plasmoids away from the disk, resulting in stronger magnetic tension from reconnected poloidal field lines in the inner region of the accretion disk. Consequently, the process of accretion is slowed down, leading to the truncation of the disk from the black hole, producing a truncated disk. Such scenarios are discussed in a later section of this chapter.

\begin{figure}
    \centering
    \includegraphics[scale=0.63]{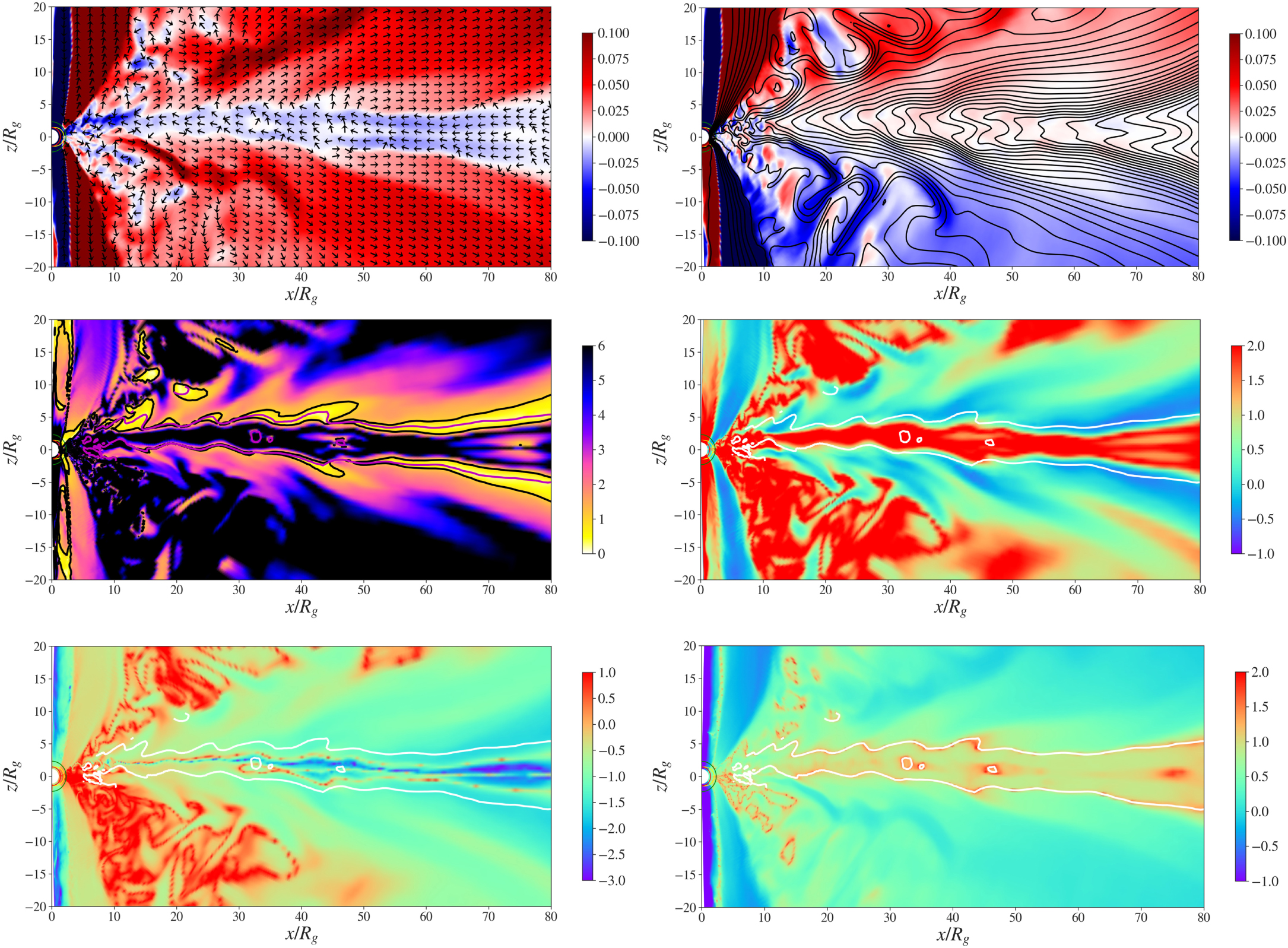}
    \caption{Snapshots at time $t=4000\,t_{\rm g}$ of the disk wind launching area for different physical variables. 
    From top left to bottom right we show the radial velocity (in color) with normalized poloidal velocity vectors (black arrows), 
    the vertical velocity (in color) with poloidal magnetic field lines (black lines), 
    the poloidal Alfv\'en Mach number (in color) superimposed with the Alfv\'en surface (black lines), 
    the plasma-$\beta$ (in color), 
    the ratio between the toroidal and poloidal magnetic field components (in color, log scale), 
    and the ratio between toroidal and poloidal velocity (color, log scale). 
    The white line (last three panels) indicate the reversal in $u_r$, thus where $u_r =0$, and where the flow changes from inflow to outflow.
    Panels taken from \cite{Vourellis-etal2019}.}
    \label{fig-vourellis}
\end{figure}

\subsection{Resistive accretion disks}
Here we discuss exemplary results of resistive GR-MHD mass loading from disk accretion into a corresponding disk wind \citep{Vourellis-etal2019}. Figure~\ref{fig-vourellis} shows a set of physical variables that characterize the transition from accretion to ejection.

We first look at the velocity distribution, in particular the radial component $u_r$. This is essentially negative in the disk (accretion) and changes to positive values above the disk (outflow). Correspondingly, the poloidal velocity vectors (normalized in the figure) turn from inward direction to outward direction (the last three panels indicate this transition with a white line). This can be considered the launching surface of the disk wind.

A similar behavior is seen in the vertical speed $u_z$ (i.e. vertical to the disk midplane), where the disk wind shows up as upward (red) and downward (blue) motion of the material above (or below, respectively) the disk surfaces. The color coding is chosen for the low speed disk wind, but note (i) the high speed BZ jet ejected from the ergosphere and (ii) the infall of material along the axis (as $B_\phi$ along the axis, no Poynting flux driven mass flow is possible here).

The disk wind becomes trans-Alfv\'enic about 5 scale heights above the midplane. Naturally, this flow is launched magnetically (see the distribution of the plasma-$\beta$, or magnetization); and, as an MHD wind, it becomes matter-dominated at some point.

The energetics of the disk outflow and respective disk jet can now be compared to the BZ-driven spine jet from the black hole. The disk wind is clearly matter-dominated, while the spine jet is Poynting flux dominated. Correspondingly, the velocity of the spine jet is highly relativistic\footnote{As for all GR-MHD simulations, the kinematics, thus the mass flux and also the speed of the BZ jet depend on the choice of the floor-density for the simulation}, while the disk jet reaches 0.1-0.2c \citep{Qian-etal2017, Qian-etal2018, Vourellis-etal2019}. However, due to the high mass flux of the disk outflow, the energy flux of the spine jet and disk jet is comparable. In some cases, the disk jet is even dominating energetically the spine jet \cite{Qian-etal2018}. The latter is obviously true in the limiting case of a non-rotating black hole that does not launch any spine jet.

\begin{figure}
    \centering
    \includegraphics[scale=0.40]{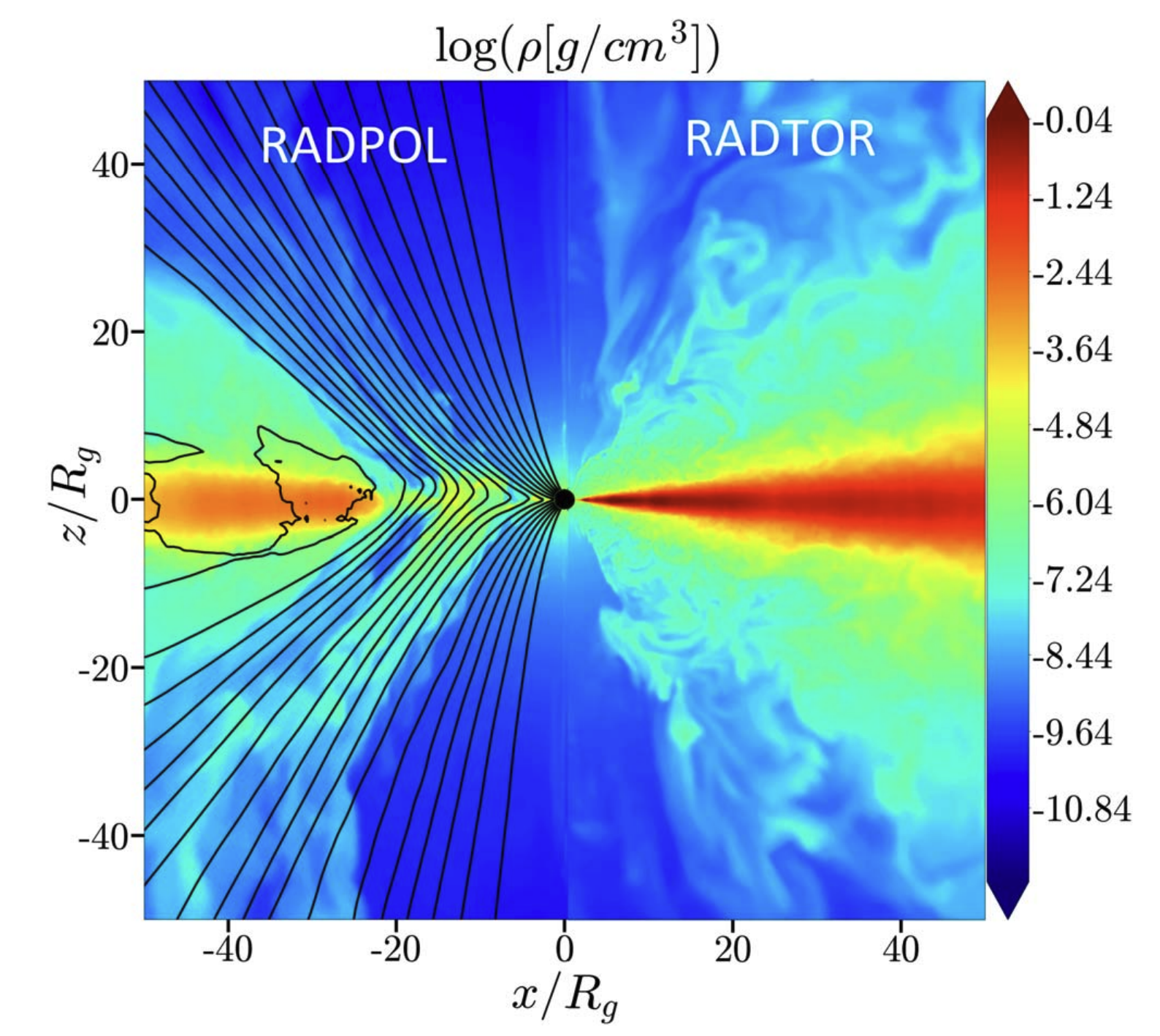}
    \caption{Density distribution and the poloidal magnetic field lines for model RADPLO and RADTOR from \cite{Liska-etal2022}.}
    \label{fig-liska}
\end{figure}

\subsection{Truncated accretion disks}
Disk truncation is thought to be responsible for the different features of BH-XRBs and AGNs, according to the standard model \citep{Esin-etal1997,Trump-etal2011,Nemmen-etal2014,Hogg-etal2018,Dihingia-etal2022a}. The truncation is thought to happen when the gas changes from a flow that is radiatively efficient and geometrically thin at the disk's edges to a flow that is radiatively inefficient and geometrically thick close to the black hole. 
Recently, Liska et. al. (2022) \cite{Liska-etal2022} with 3D simulations  of thin disks have demonstrated the formation of a truncated accretion disk due to the presence of a strong poloidal magnetic flux. In Fig.~\ref{fig-liska}, we show the density distribution of the two models they studied. 
In model RADTOR, there is no net poloidal magnetic flux. 
In model RADPOL, on the other hand, there is only poloidal magnetic flux, and it reaches a MAD state after time evolution. 
They show that when the magnetic flux in the model RADPOL's inner disk reaches its maximum, there is a sharp change between an outer disk supported by radiation pressure and an inner corona supported by magnetic pressure. For their model, the truncation radius is found to be $r_{\rm tr}=20r_g$. On the other hand, we do not observe such truncation in the RADTOR model (see Fig.~\ref{fig-liska} for reference).

\begin{figure}
    \centering
    \includegraphics[scale=0.30]{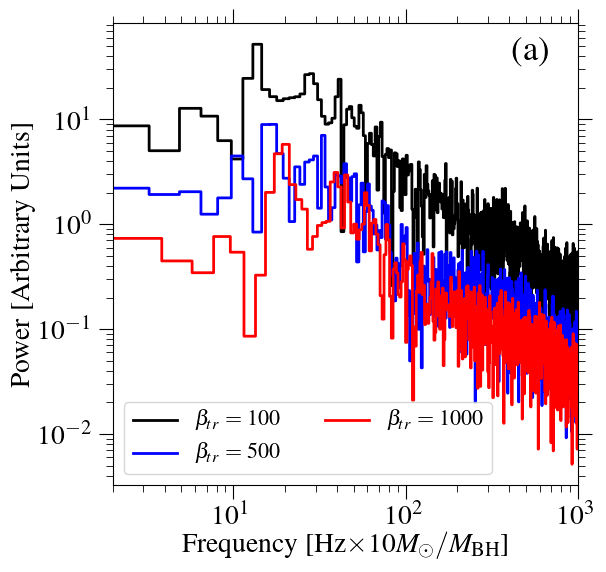}
    \includegraphics[scale=0.30]{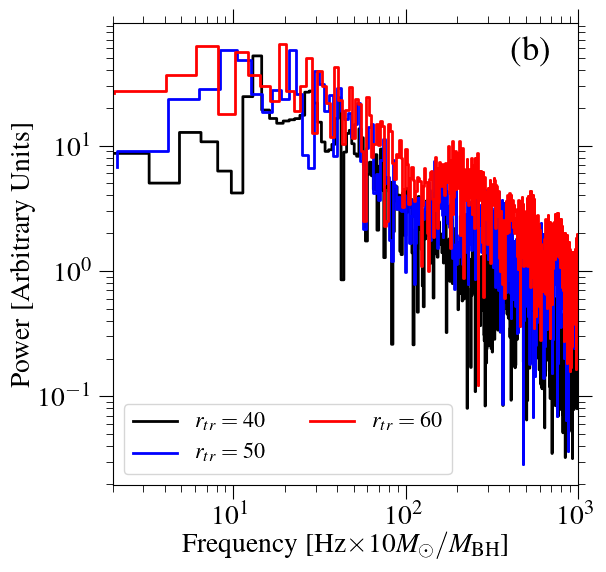}
    \caption{PDS corresponding to the accretion rate profile for different simulation models with different magnetic field strengths and truncation radius from \cite{Dihingia-etal2022a}.}
    \label{fig-pds}
\end{figure}

More interesting features of the truncated accretion disk have been reported in Dihingia et. al. 2022 \cite{Dihingia-etal2022a}. According to them, the reconnection events and the plasmoids that form because of active tearing mode instabilities are very important to the accretion flow dynamics of the truncated disk. They make poloidal magnetic fields vertical to the equatorial plane, and that creates a strong magnetic tension force in the midplane of the disk. The accretion flow is temporarily stopped by this force's resistance. Over time, the ram pressure of the flow in the inner edge becomes higher against this force because of the transport of angular momentum outwards (because of MRI and disk winds), which causes accretion to happen again. As the simulation goes on, this process keeps going as the inner accretion flow oscillates. They show that the flow inside the truncation radius oscillates or the inner radius oscillates in a quasi-periodic manner. The frequency of oscillations depends on the magnetic field strengths, truncation radius, and resistivity of the magnetized flow. We reproduced the power density spectra (PDS) of accretion rate at the horizon from their paper in Fig.~\ref{fig-pds}. 

In contrast to ideal MHD models, no formation of plasmoids is observed with the increase in resistivity. Additionally, the turbulent characteristics of the inner accretion flow are subdued as magnetic resistivity rises. The Kelvin-Helmholtz instability (KHI) across the inflow and outflow boundaries experiences suppression due to increased magnetic resistivity. Moreover, the frequency of oscillation rises with the increase in magnetic resistivity. Furthermore, the radius at which the balancing magnetic tension force and the ram pressure occur decreases proportionally with the escalation of magnetic resistivity (see \cite{Dihingia-etal2022a}). Further evidence of transient shocks during the highly accreting stage is uncovered during oscillation. These transient shocks manifest as an extended hot post-shock corona surrounding the black hole, exerting an influence on its radiative properties. Radiative transfer calculations reveal diskernible signatures of these oscillations through modulation in the edge-brightened structure of the accretion disk. These modulations strongly relate to the observed quasi-periodic oscillations (QPOs) in BH-XRBs during their outburst \citep {Belloni-etal2011}.

\subsection{High-soft to low-hard transition}
It has been well known for many decades that a geometrically thin disk can suitably describe the high-soft state of an outburst. Similarly, geometrically thick flow close to the black hole can describe low-hard state of an outburst \citep{Narayan-Yi1995,Chakrabarti-Titarchuk1995,Esin-etal1997,Narayan-McClintock2012,Kylafis-Belloni2015,Kylafis-etal2018,Chakrabarti2018}. Recent, GR-MHD simulations also suggest that geometrically thick hot magnetised flow could be a reasonable choice for the hard state of BH-XRBs \citep[e.g.,][]{Dexter-etal2021}. Although there is doubt if the flow in the hard state is MAD or not \cite[e.g.,][]{You-etal2023,Fragile-etal2023}. Still, no one knows how or why the different states change or how the different states are connected together. Accrution flow simulations that look like the real thing might help us understand the changes in states. 
Still, simulating an outburst over a realistic time range (hours to tens of days) is hard, especially when using GR-MHD. 

\begin{figure}
    \centering
    \includegraphics[scale=0.35]{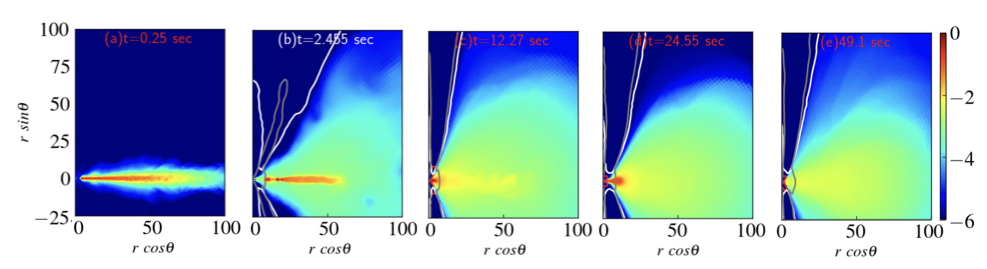}
    \caption{Logarithmic time-average density distribution at different times marked on the top of the figures. The time averaging is performed within $0.1$ sec around the marked time. The white and green solid lines show $-hu_t=1$ and $\sigma=1$ contours, respectively. The panels are taken from \cite{Dihingia-etal2023}.}
    \label{fig-hslh}
\end{figure}

Recently, Dihingia et. al. (2023) \cite{Dihingia-etal2023} investigated the decaying phase of an outburst in BH-XRBs. 
For this setup, an initial thin disk is considered near the black hole, and low-angular matter is injected far from the black hole. The density evolution of their simulation is reproduced in Fig.~\ref{fig-hslh}. These simulations include radiative cooling following \cite{Dihingia-etal2022,Dihingia-etal2023a}.
Initially, the injected matter interacts with the thin disk and forms a TCAF. The thin disk at the equatorial plane is where the majority of the radiation in such a flow originates. However, some hard X-rays are expected to be observed by the inverse Compton process because of the hot corona surrounding the thin disk. The soft-intermediate phases of an outburst could therefore be explained by such a TCAF \citep[][and references therein]{Chakrabarti-Titarchuk1995,Chakrabarti2018}. Over time, more matter is injected, leading to the transformation of the accretion flow into a geometrically thick and hot torus. The size of the torus diminishes, leaving behind very low-density hot matter surrounding the BH. Such a flow structure resembles a radiatively inefficient quiescent state (RIAF) \citep[e.g.,][]{Narayan-etal1996,Narayan-etal1997,Ferreira-etal2006}. Thus, the simulation shows a progression from a soft state to a soft-intermediate state, then to a hard-intermediate state, followed by a hard state, and finally to a quiescent state. The detection of jet properties aligns with expectations from many observations, indicating the presence of jets only in intermediate states. A distinct variation in QPO frequency is observed throughout the evolution. On an equal decaying time scale for mass and angular momentum, the accretion rate at different radii decays with time. This feature is expected in the decaying phase of commonly known outbursts in BH-XRBs \citep{Esin-etal1997,Ferreira-etal2006,Kylafis-Belloni2015}. The high-resolution run shows similar results for accretion rates and QPO frequencies, but a higher accumulation of magnetic flux around the event horizon results in stronger jets.

\begin{figure}
    \centering
    \includegraphics[scale=0.63]{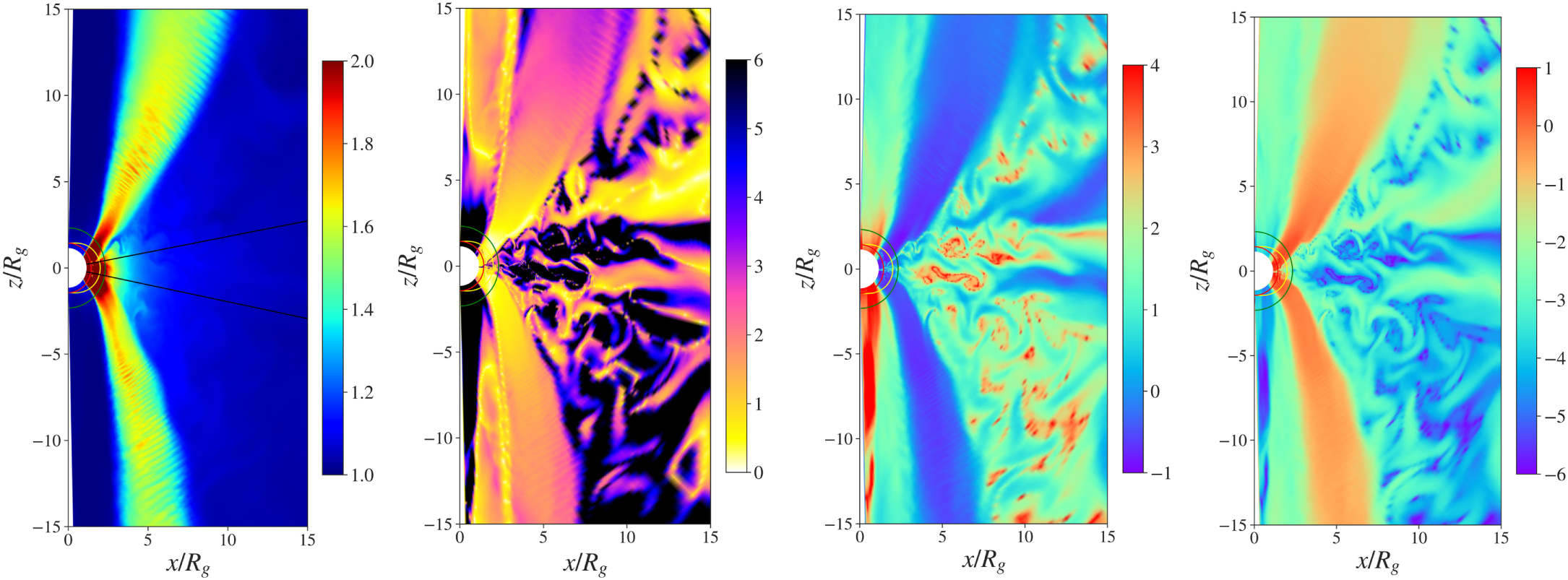}
    \caption{Snapshots at time $t=4000\,t_{\rm g}$ of magnetohydrodynamic accretion–ejection structure close to the black hole, $r < 15$. 
    Shown is the Lorentz factor (left), the poloidal Alfv\'en Mach number (log scale; second from left), the plasma-$\beta$ (log scale; second from right), 
    and the magnetization (log scale; right). The high Alfv\'en Mach number indicates flows that are dominated by kinetic energy,
    regardless of the highly magnetized area. The red semicircle marks the horizon, the yellow line marks the ergosphere, and the green 
    line indicates the radius of the ISCO.
    Panels taken from \cite{Vourellis-etal2019}.}
    \label{fig-reconn}
\end{figure}

\begin{figure}
    \centering
    \includegraphics[scale=0.85]{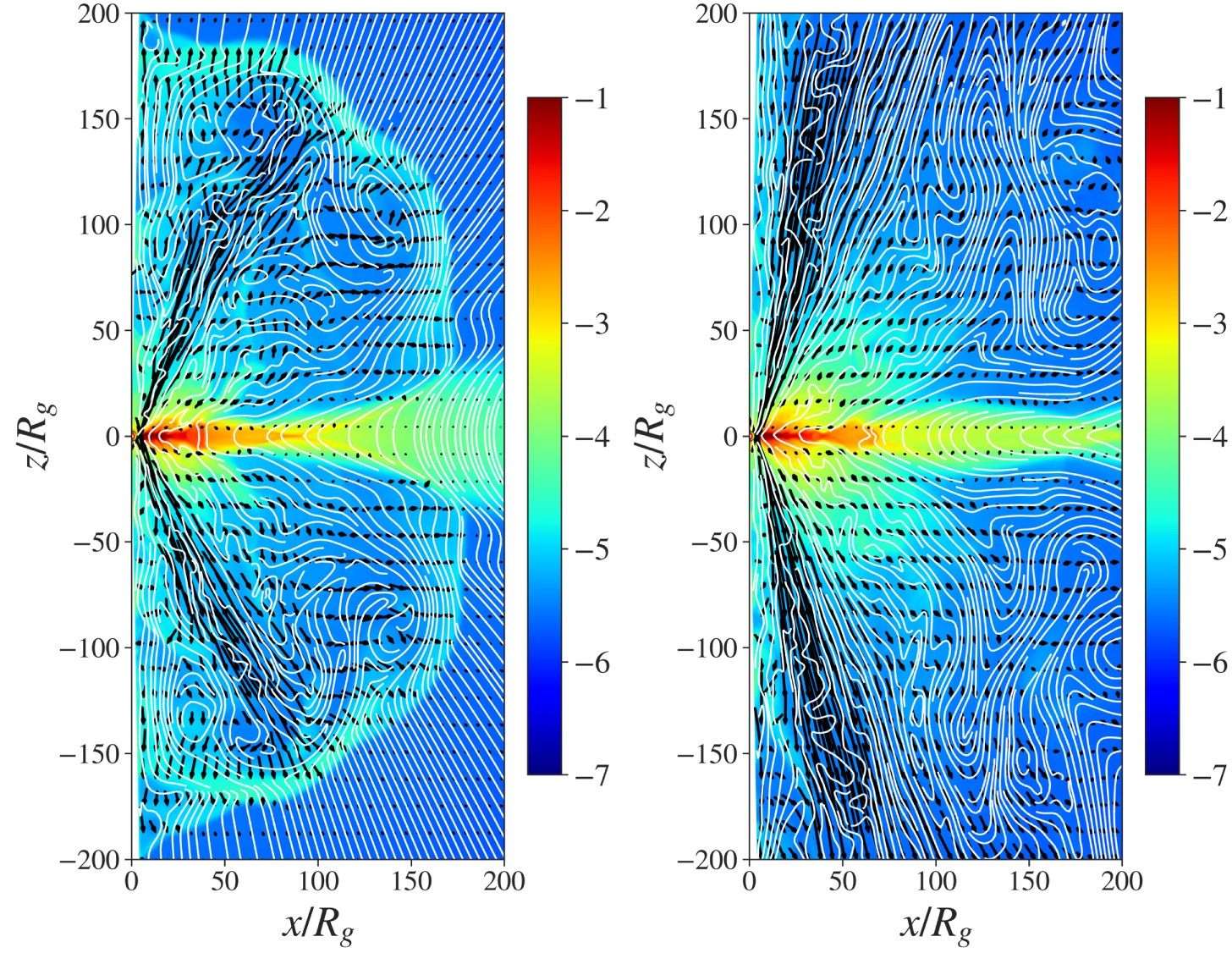}
    \caption{Snapshots at time $t=2000, 4000\,t_{\rm g}$ (from left to right) of magnetohydrodynamic accretion–ejection structure enclosing a BZ jet
     at larger distanced from th elaunching reqion. 
     Shown is the density distribution (in log scale) and the poloidal magnetic field lines (white lines), while the poloidal velocity field 
     is represented by the black arrows.
    Panels taken from \cite{Vourellis-etal2019}.}
    \label{fig-reconn2}
\end{figure}

\subsection{Magnetospheric reconnection and ejection of plasmoids}
Resistive MHD allows for physical reconnection of the magnetic field. This process is of particular importance for the magnetic field close to the black hole, where we typically have a strong anti-aligned magnetic field along the equatorial plane. Here, magnetic reconnection works very efficiently, effectively disturbing the smooth inflow-outflow close to the inner disk radius. As a result, magnetic islands and plasmoids are ejected along the transition region between the inner BZ jet and the surrounding disk wind.

Figure~\ref{fig-reconn} visualizes the processes involved. The high speed BP jet is visible predominantly in the Lorentz factor. This jet starts sub-Alfv\'enic and becomes super-Alfv\'enic further downstream. We can see the turbulent flow region that the reconnection along the mid-plane in the plunge region towards the black hole has perturbed right outside this jet. This region is not magnetically dominated, with rather high plasma-$\beta$, respectively, low magnetization.

In our simulations, these plasmoids are rather long-lived, as resistivity decreases with altitude. When these magnetic islands reach the ideal MHD regime at sufficiently large distances from the black hole and the disk, they may reconnect only by numerical resistivity. In our case, they are just ejected out of the simulation domain. However, on their way downstream, they also perturb the (light) BZ jet (see Figure~\ref{fig-reconn2}). Plasmoid generation and ejection have been investigated in more detail and have been confirmed by \cite{Nathanail-etal2020,Nathanail-etal2022}.

\begin{figure}
    \centering
    \includegraphics[scale=0.6]{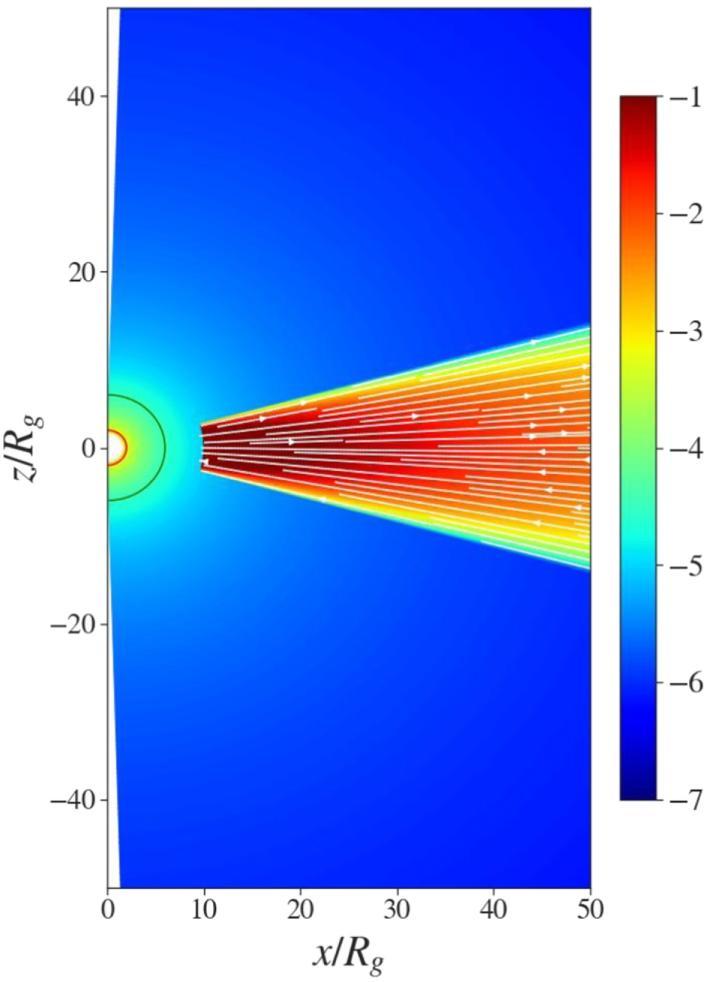} 
    \includegraphics[scale=0.6]{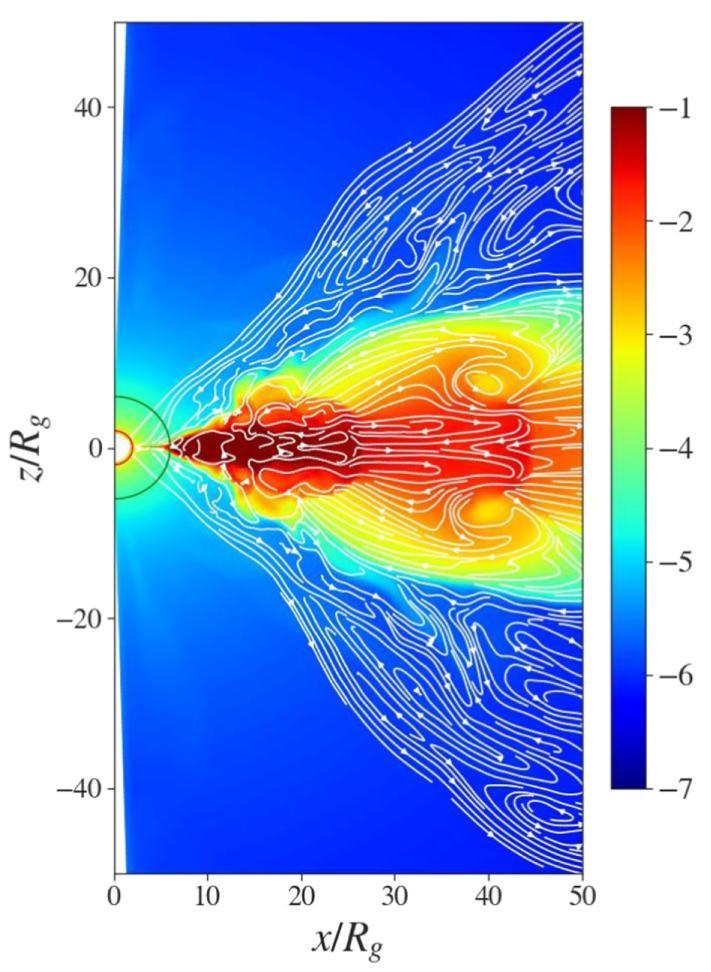}
    \includegraphics[scale=0.6]{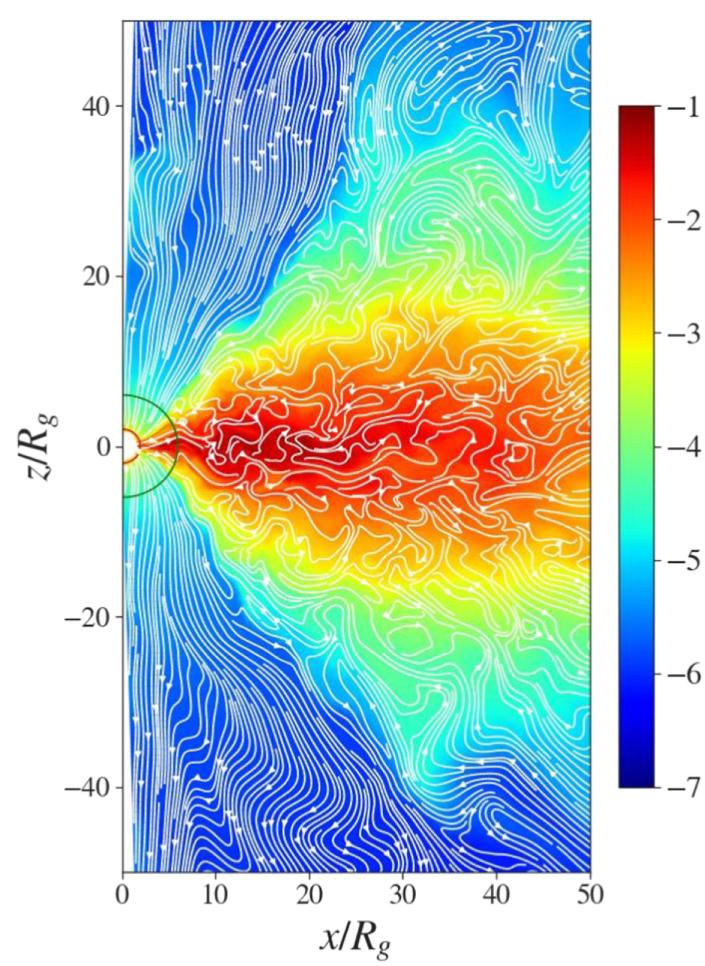}
    \caption{Evolution of density and the poloidal magnetic field (white lines) over time.
    Left is the initial setup with a purely radial (weak) seed field.
    The middle panel show the situation at $t=3000\,t_{\rm g}$, the right panel at $t=12000\,t_{\rm g}$. 
    The red line around the (0, 0) point is the event horizon, and the green line denotes the radius of the ISCO.
    It is clearly seen how the field evolves, expands to larger radii, and becomes advected to the black hole.
    Note that the the field lines plotted indicate the field structure, but not the field strength.
    The latter increase by a factor 10,000 over the run time of this simulation. 
    Panels taken from \cite{VourellisFendt2021}.}
    \label{fig-dynamo-glob}
\end{figure}

\begin{figure}
    \centering
    \includegraphics[scale=0.85]{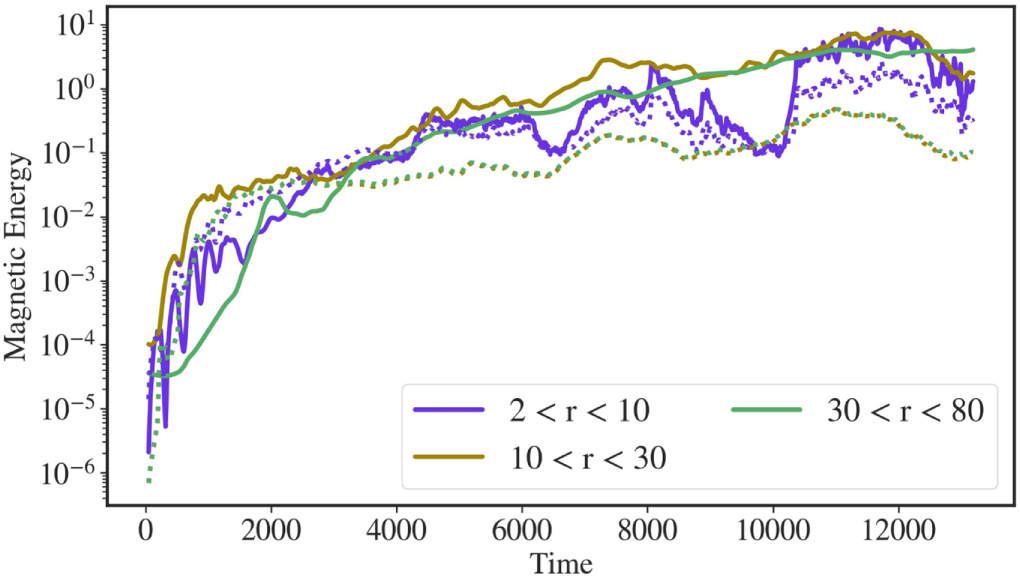} 
    \caption{Evolution magnetic field strength during the dynamo process.
    Shown is the averaged disk magnetic energy over time.
    Different disk areas are distinguished, such as $2<r<10$, $10<r<30$, and $30<r<80$
    (radii $r$ normalized to $R_{\rm g}$). 
    Panels taken from \cite{VourellisFendt2021}.}
    \label{fig-evolv-disk}
\end{figure}

\begin{figure}
    \centering
    \includegraphics[scale=0.32]{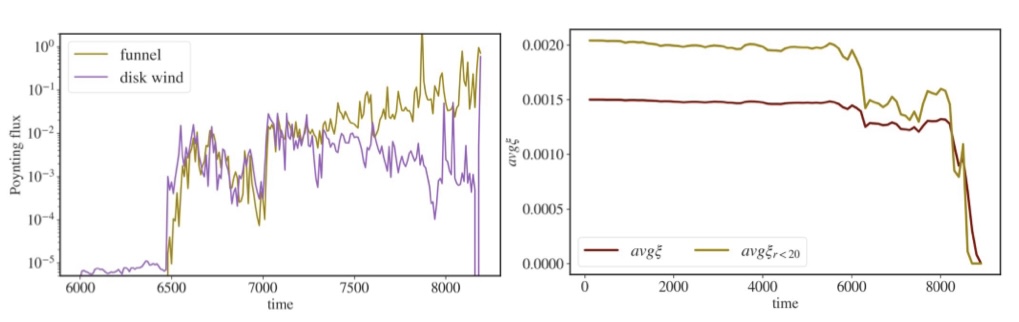}
    \caption{Evolution magnetic field strength during the dynamo process.
    Shown is the Poynting flux for the funnel BZ jet, and the disk wind (left).
    The flux is integrated along a sphere with radius $r=35\,R_{\rm g}$, while the BZ considered 
    for the segment $0\deg <\theta < 25\deg$ and the disk wind for the segment 
    $25\deg <\theta < 65\deg$. The values are averaged over the corresponding segments at the
    lower hemisphere.
    The corresponding evolution of the absolute dynamo parameter $\xi$ is shown at the bottom, 
    averaged in space for the whole disk, and for the inner disk area, $r < 20\,R_{\rm g}$.
    Panels taken from \cite{VourellisFendt2021}.}
    \label{fig-evolv-poynting}
\end{figure}

\subsection{A GR-MHD disk dynamo-generated disk jet}
We subsequently discuss results considering the action of a mean-field GR-MHD disk dynamo. As found in previous studies for non-relativistic disks, the GR-MHD disk dynamo is also able to generate a strong poloidal magnetic flux anchored in the disk. This magnetic field can (i) drive a disk wind that potentially collimates into a disk jet and can also (ii) be advected to the central spinning black hole, thereby generating a strong BZ jet.

Thus, our approach self-consistently generates a two-component jet structure from a disk seed field. In comparison to the non-relativistic studies, which result in a smooth disk-jet structure after about $100,000$ inner disk revolutions, the magnetic field of the GR-MHD dynamo is much more disturbed due to the strong reconnection going on in the plunge region and the very inner disk. This is shown in Figure~\ref{fig-dynamo-glob} with the time evolution of the global magnetic field structure. The simulation starts with a weak, purely radial seed field that is confined to the disk. At the final state, a global magnetic field has evolved, in particular a strong field that has been advected to the black hole ergosphere and drives a strong BZ jet, in addition to a disk wind.

The increase in magnetic field strength is shown in Figure~\ref{fig-evolv-disk} and \ref{fig-evolv-poynting}. Figure~\ref{fig-evolv-disk} shows the evolution of the magnetic energy in the disk for selected disk areas. We clearly see an exponential increase with a flattening due to dynamo quenching, resulting in a $10^6$ increase of the disk magnetic energy (for this choice of dynamo strength $\xi$ and dynamo quenching). We want to stress the point that jet launching, in particular by the BP process, requires a large global magnetic flux. In this respect, plotting the magnetic energy or the absolute field strength could be misleading, as for the global magnetic flux, the small-scale field components may actually cancel out.

It is therefore interesting to look at the Poynting flux of the two-component jet structure. This is shown in Figure~\ref{fig-evolv-poynting}, where we see how the Poynting flux evolves in time. The flux is measured along a sphere around the black hole with a radius of $35\,R_g$. Note that it takes some time ($t=6000\,t_{\rm g}$) until the flux generated in the disk has reached that distance. After that, the flux increases with similar strength for both components: the inner funnel flow, i.e., the BZ jet, and the outer disk wind. After a further period of time ($t\simeq 8000\,t_{\rm g}$) the disk dynamo is quenched as the disk magnetic field has reached a (chosen) equipartition value. Thus, the disk Poynting flux levels off, even slightly decreasing. On the other hand, the advecting of magnetic flux to the black hole is still going on, and the Poynting flux of the funnel jet still increases. That increase is actually on the costs of the disk flux, as it is advected away from the disk area towards the black hole.

Corresponding to the magnetic flux evolution, the lower panel of Figure~\ref{fig-evolv-poynting} shows the evolution of the dynamo parameter $\xi$. The onset of dynamo quenching is simply seen as a decrease of $\xi$.

\section{Summary and outlook}
Over the past few decades, GR-MHD simulations have been performed for different astrophysical scenarios, including thin disk configurations. In this chapter, we have discussed some of the recent results considering highly resolved GR-MHD simulations of thin disks. In particular, we have discussed non-ideal MHD effects such as resisitivity, reconnection, and a mean-field dynamo.

Future work should be devoted to high-resolution 3D simulations of the inner disk area in order to investigate variable accretion events and the time scales of the same. Also, as thin disks are thought to be cold and radiative, radiation modeling of the disk material seems essential. This is because of two reasons (at least): one is the connection and the comparison to observations, in particular the opportunity of fitting distinct sources. The other is the scaling of mass fluxes, as in pure MHD modeling the disk mass is unconstrained (as in the simple Keplerian problem). Radiation modeling in fact requires applying physical densities, which can constrain mass accretion and ejection rates. Essential steps in this direction have already been performed \cite{Noble-etal2009, McKinney-etal2014,Liska-etal2022, Liska-etal2023}, mostly not for the thin disk ansatz, but also a few for thin disks \cite{Noble-etal2011,Morales-etal2018}, none for resistive disks.

In this book chapter, we show that an initial thin disk could explain the decaying stage of an outburst with GR-MHD simulations. It should now be a priority to simulate the full cycle of an outburst (including the rising phase). Note that a realistic outburst in BH-XRBs could be a few hours to months long and rarely even years  \cite[for e.g.,][]{Yan-Yu2015}, which is impossible to perform under the current framework. Therefore, more quantitative and parameter space surveys are required to correlate the realistic decay time with the physical properties of injected matter, the initial thin accretion disk, or the properties of the central BH ($a, M_{\rm BH}$, etc.). 

Most of the studies do not include self-consistent radiative cooling processes that may be present in the flow. The radiative cooling may not be important for lower accretion rates ($\dot{M}<10^{-7}$ Eddington units, \cite{Yoon-etal2020,Dihingia-etal2022}). Also, the radiative calculations require the electron temperature. Throughout the different studies we discussed here, a single-temperature approximation has been employed. Such studies usually rely on temperature scaling relations, e.g., $R-\beta$ model \cite{Moscibrodzka-etal2016}. Such relations may result in an overestimation of electron temperatures and are hence not recommended \citep[e.g.,][]{Sadowski-etal2017,Mizuno-etal2021,Dihingia-etal2022}. Hence, the adoption of a robust two-temperature approach is recommended for achieving self-consistent results.

After all the efforts in the field, we still don't know the mass composition of relativistic jets or the very physical mechanism that leads to the ejection of the observed radio knots. In interpreting the highly relativistic jet region from GR-MHD simulations, one always needs to take prior caution. This region is filled with floor quantities and is therefore not perfectly reliable \citep[e.g.,][]{McKinney-etal2012,Porth-etal2017}. As a result, GR-MHD alone cannot explain the fundamental physical mechanisms behind the launch of highly relativistic jets. General relativistic perticle-in-cell (GR-PIC) simulations may be needed, which are currently under development \citep[e.g.,][]{Levinson-Cerutti2018,Bacchini-etal2019,Kisaka-etal2020,Hirotani-etal2023}. These simulations may provide a more accurate representation of the complex processes involved in the formation of relativistic jets. By combining GR-MHD and GR-PIC simulations, scientists may be able to gain a more comprehensive understanding of the physics driving these phenomena. 

\section*{Acknowledgements}
C.F. is grateful to Scott Noble for permission to use the original HARM3D code for further development (resistivity, dynamo).
C.F. acknowledges fundamental contributions during code development by PhD students Qian Qian and Christos Vourellis, 
as well as critical help during the implementation of resistivity in HARM3D by Matteo Bugli and Scott Noble.
C.F. acknowledges support by the German Research Foundation DFG via the research unit FORS\,5195. 
Simulations co-authored by C.F. were performed on the ISAAC and VERA clusters of the Max Planck Institute for Astronomy. 
I.K.D. acknowledges the support from the National Key R\&D Program of China (No. 2023YFE0101200), the National Natural Science Foundation of China (Grant No. 12273022), and the Shanghai Municipality orientation program of Basic Research for International Scientists (grant no. 22JC1410600), and Max Planck partner group award at Indian Institute Technology of Indore, India. Simulations co-authored by I.K.D. were performed on the TDLI-Astro cluster in Tsung-Dao Lee Institute, Pi2.0, Siyuan Mark-I clusters in the High-Performance Computing Center at Shanghai Jiao Tong University, and Max Planck Gesellschaft (MPG) super-computing resources.


\end{document}